\documentclass[preprint]{aastex}
\newcommand{\msun}{{\rm M_\odot}}
\newcommand{\mjup}{{\rm M_J}}
\newcommand{\rsun}{R_\odot}
\newcommand{\kms}{km\,s$^{-1}$}

\shorttitle{The dynamical evolution of low-mass systems formed through disc fragmentation}
\shortauthors{Li, Kouwenhoven, Stamatellos \& Goodwin}

\begin{document}

\title{The dynamical evolution of low-mass hydrogen-burning stars, brown dwarfs and planetary-mass objects formed through disc fragmentation}

\author{Yun Li\altaffilmark{1}}
\affil{Department of Astronomy, School of Physics, Peking University, Yiheyuan Lu 5, Haidian Qu, Beijing 100871, China}
\affil{Kavli Institute for Astronomy and Astrophysics, Peking University, Yiheyuan Lu 5, Haidian Qu, Beijing 100871, China}
\email{yunli@pku.edu.cn}

\author{M.B.N. Kouwenhoven\altaffilmark{2}}
\affil{Kavli Institute for Astronomy and Astrophysics, Peking University, Yiheyuan Lu 5, Haidian Qu, Beijing 100871, China}
\affil{Department of Astronomy, School of Physics, Peking University, Yiheyuan Lu 5, Haidian Qu, Beijing 100871, China}

\author{D. Stamatellos\altaffilmark{3}}
\affil{Jeremiah Horrocks Institute for Mathematics, Physics \& Astronomy, University of Central Lancashire, Preston, PR1\,2HE, UK}

\and

\author{S.P. Goodwin\altaffilmark{4}}
\affil{Department of Physics \& Astronomy, The University of Sheffield, Hicks Building, Hounsfield Road, Sheffield S3\,7RH, UK}

\begin{abstract}
Theory and simulations suggest that it is possible to form low-mass hydrogen-burning stars, brown dwarfs and planetary-mass objects via disc fragmentation. As disc fragmentation results in the formation of several bodies at comparable distances to the host star, their orbits are generally unstable. Here, we study the dynamical evolution of these objects. We set up the initial conditions based on the outcomes of the smoothed-particle hydrodynamics (SPH) simulations of \cite{stamatellos2009a}, and for comparison we also study the evolution of systems resulting from lower-mass fragmenting discs. We refer to these two sets of simulations as set~1 and set~2, respectively. At 10~Myr, approximately half of the host stars have one companion left, and approximately 22\% (set~1) to 9.8\% (set~2) of the host stars are single. Systems with multiple secondaries in relatively stable configurations are common (about 30\% and 44\%, respectively). The majority of the companions are ejected within 1~Myr with velocities mostly below 5~\kms, with some runaway escapers with velocities over 30~\kms. Roughly 6\% (set~1) and 2\% (set~2) of the companions pair up into very low-mass binary systems, resulting in respective binary fractions of 3.2\% and 1.2\%. 
The majority of these pairs escape as very low-mass binaries, while others remain bound to the host star in hierarchical configurations (often with retrograde inner orbits).
Physical collisions with the host star (0.43 and 0.18 events per host star for set~1 and set~2, respectively) and between companions (0.08 and 0.04 events per host star for set~1 and set~2, respectively) are relatively common and their frequency increases with increasing disc mass. Our study predicts observable properties of very low-mass binaries, low-mass hierarchical systems, the brown dwarf desert, and free-floating brown dwarfs and planetary-mass objects in and near young stellar groupings, which can be used to distinguish between different formation scenarios of very low-mass stars, brown dwarfs and planetary-mass objects.
\end{abstract}

\keywords{(stars:) brown dwarfs -- stars: formation -- stars: kinematics and dynamics -- stars: low-mass -- (stars:) planetary systems}

%%%%%%%%%%%%%%%%%%%%%%%%%%%%%%%%%%%%%%%%%%%%%%%%%%%%%%%%%%%%%%%%%%%%%%%%%%%%%
%%%%%%%%%%%%%%%%%%%%%%%%%%%%%%%%%%%%%%%%%%%%%%%%%%%%%%%%%%%%%%%%%%%%%%%%%%%%%
%%%%%%%%%%%%%%%%%%%%%%%%%%%%%%%%%%%%%%%%%%%%%%%%%%%%%%%%%%%%%%%%%%%%%%%%%%%%%

\section{Introduction}

The Galactic field contains a large number of brown dwarfs and stars below $0.2~\msun$ \citep{kroupa2001, chabrier2003, chabrier2005}. The low-mass regime of the "stellar" initial mass function (IMF) is occupied by three types of objects: low-mass hydrogen-burning stars (LMSs), which have enough mass to sustain hydrogen burning in their cores ($>80~\mjup$, where ${\rm M_J}$ is the mass of Jupiter), brown dwarfs (BDs), which are not massive enough to sustain hydrogen burning but they can sustain deuterium burning ($\sim13-80~\mjup$), and  planetary-mass objects (PMOs) that can not burn deuterium ($<13~\mjup$). This classification is based solely on the mass of an object but the possibility that all these types of objects can form by the same mechanism can not be excluded \citep[see, e.g.,][]{whitworth2007, luhman2012}.

These low-mass objects may form similarly to higher-mass stars (e.g. Sun-like stars) either by the collapse of pre-(sub)stellar cores \citep[e.g.,][]{andre2014, lomax2015} or by gravitational instabilities in protostellar discs. Disc fragmentation reproduces the critical constraints set by the observed statistical properties of low-mass objects, i.e. the shape of the mass distribution of low-mass objects \citep{thies2007}, the differences in binary statistics with respect to higher-mass stars \citep[][]{thies2008, dieterich2012}, the lack of BDs as close companions to Sun-like stars \citep[the BD desert; see, e.g.,][]{marcy2000, grether2006, dieterich2012, evans2012}, the presence of discs around BDs, and also the statistics of low-mass binary systems, and the formation of free-floating PMOs \citep[e.g.,][]{stamatellos2007b, stamatellos2009a}. In particular, multiple (rather than just one or two) fragments per pre-stellar core can also explain the observed properties of the IMF and the primordial binary population \citep{goodwin2008, goodwin2009, holman2013}. The resulting few-body systems are generally very unstable and decay rapidly, which is consistent with observations indicating that the multiplicity fraction for different populations decreases with age \citep[see, e.g.,][for a review]{reipurth2014}.

Numerical studies of disc fragmentation \citep[e.g.][]{stamatellos2009a,tsukamoto2013} are not statistically robust  as the high computational cost to treat gas thermodynamics allows only for a limited number of simulations to be performed. To improve on this we combine a small number of gas simulations with a large number of subsequent $N$-body simulations.  
We assume that discs around solar-type stars fragment to produce a few low-mass objects (with initial orbital properties that are provided by the SPH simulations), and then we follow the long-term dynamical evolution of these systems using pure $N$-body techniques, assuming that most of the gas in the disc has been converted into low-mass objects or accreted onto the host star \citep[see][for a discussion on disc dispersal]{alexander2014}. This assumption allows us to perform a large ensemble of realisations of systems formed by disc fragmentation and produce statistically robust results. Our aim is to determine whether the observed properties of LMSs, BDs, and PMOs (orbital properties around their parent stars, the BD desert, binary properties among low-mass stars, and properties of free-floating BDs and PMOs) are consistent with the model of disc fragmentation.

Similar studies exploring the formation of binary stars during the dynamical decay of few-body systems have been performed in the past but for higher mass stars and/or for different initial configurations. 
\cite{mcdonald1993} and \cite{clarke1996}, for example, study the decay of small-$N$ systems, picking masses randomly from the IMF.  In their study binaries form by dynamical interactions as the system decays. They are able to reproduce the observed binary fraction and mass ratio distribution if dissipative encounters (due to gas) are taken into account.
\cite{sterzik1998} examine the dynamical decay of few-body ($N=3-5$) systems formed by fragmenting clouds cores  (not discs as in this paper) for which they arbitrarily set the parameters. They are able  to reproduce well the mass ratio distribution for binary M-type stars, and also their semi-major axis distribution. However, the stellar masses that they consider are all above $0.2~\msun$.  \cite{durisen2001} follow a two-step approach in which they use a core mass function (CMF) to compute the total mass of the bodies (assumed to form by cloud core fragmentation), and then they pick up the masses of the stars/BDs using an IMF.  This approach produces binary properties that are in better agreement with observations  than the  one-step approach, i.e. picking random masses directly from the IMF. \cite{sterzik2003} have extended this study to the brown dwarf regime by sampling the core mass function to produce an ensemble of cores and the initial mass function to get the masses of the stars/BDs for their $N$-body simulations. Their models result in populations that are consistent with the observed BD binary fraction and corresponding mass ratio distribution.

\cite{hubber2005} investigate binary star formation, assuming that fragmentation happens in a ring that gives rise to a small ($N<6$) cluster that  decays dynamically. They follow the dynamical decay through $N$-body simulations. For a set of parameters they are able to reproduce the binary fraction for low-mass objects but they do not discuss  the properties of low-mass binaries. Systems formed through disc fragmentation generally result in unstable systems where strong gravitational scattering is common, often resulting in ejections.
\cite{umbreit2005} investigate the ejection scenario for brown dwarf formation and the associated pairing at low masses using $N$-body simulations. They follow the evolution of triple BD proto-systems (thought to form in a small cloud core, not in discs as in this paper). All three BDs have masses of $0.05~\msun$. They assume that the masses of the BDs increase with time  using a simple prescription for accretion and assuming a constant accretion rate. Some of the BDs in their simulations grow in mass to become LMSs. They reproduce rather well the observed semi-major axis distribution for BDs.
The recent work of \cite{forgan2015} discusses the evolution of disc-fragmented systems in clustered stellar environments. Based on the population synthesis models of \cite{forgan2013} they generate systems with an average of 2.3~fragments with initial semi-major axes of approximately $1-100$~AU, and do not model physical collisions. Due to the relatively small number of companions, they are able to directly model the evolution of the companions in star clusters, in which close stellar encounters may enhance the disruption of disc-fragmented systems.

In this study we follow a similar approach in which we evolve the systems formed by disc fragmentation and examine their properties after they have obtained a stable configuration. The main difference with previous studies is the initial configuration of the system: we assume that a few low-mass objects (LMSs, BDs, PMOs  with masses up to $200~\mjup$) form by disc fragmentation and therefore their orbits are initially almost on the same plane, and they orbit their host star, which has a much larger mass ($0.7~\msun$). Our goal is to examine the orbital properties of these low-mass objects (e.g., distances from their host stars, eccentricities, orbital planes) and their binary properties (e.g., semi-major axes, mass ratios, eccentricities) and compare them with the observed properties of LMSs, BDs and PMOs.

The computational method and initial conditions are described in Section~\ref{section:method} and~\ref{section:initial}, respectively. The decay of the systems is discussed in Section~\ref{section:decay}, and the properties of the physical collisions and escapers, the evolution of the orbital elements, and the formation of binaries and higher-order multiple systems in Section~\ref{section:other}. Finally, we interpret our results in Section~\ref{section:discussion} and draw our conclusions in Section~\ref{section:conclusions}.

%%%%%%%%%%%%%%%%%%%%%%%%%%%%%%%%%%%%%%%%%%%%%%%%%%%%%%%%%%%%%%%%%%%%%%%%%%%%%
%%%%%%%%%%%%%%%%%%%%%%%%%%%%%%%%%%%%%%%%%%%%%%%%%%%%%%%%%%%%%%%%%%%%%%%%%%%%%
%%%%%%%%%%%%%%%%%%%%%%%%%%%%%%%%%%%%%%%%%%%%%%%%%%%%%%%%%%%%%%%%%%%%%%%%%%%%%

\section{Methodology} \label{section:method}

We assume that solar-type stars form with discs that fragment to produce several low-mass objects. We use the results of previous SPH (gas) simulations of disc fragmentation \citep{stamatellos2009a} to determine the initial properties of the low-mass objects that form in the disc and  then we follow their long-term dynamical evolution using pure $N$-body techniques. We therefore ignore any residual disc gas in our simulations, assuming that this has been accreted onto the host star or converted into low-mass objects. This allows us to simulate the evolution of a large number of systems which is not computationally feasible when gas is accounted for. 

The $N$-body simulations are carried out with the MERCURY6 software package \citep{chambers1999}, using the general Bulirsch-Stoer algorithm \citep{stoer1980, press2002}. This is an accurate but slow algorithm that conserves the fractional energy  well in most runs (with $\Delta E/E$ between $10^{-8}$ and $10^{-5}$). Several runs, however (among which many stable triple or quadruple configurations), result in fractional energy errors $\Delta E/E > 10^{-3}$. These are re-run using the RADAU algorithm \citep{everhart1985} to achieve an equally small energy conservation. Although the RADAU algorithm is unable to handle extremely close encounters and very eccentric orbits, it accurately evolves the systems that resulted in large fractional energy errors when using the Bulirsch-Stoer algorithm. The energy conservation of the combined set of simulations thus remains below $10^{-3}$, with the vast majority of the systems having $\Delta E/E$ in the range $10^{-8}-10^{-5}$.

We refer to the central star in each system (i.e. the star whose disc has fragmented) as the {\it host star}, and the products of disc fragmentation as the {\it secondary objects} (or {\em secondaries}). As the systems dynamically evolve, some of the secondary objects remain bound to the host star while others escape. Secondary objects can form a bound binary system that orbits the host star in a triple configuration, or escapes from the host system as a low-mass binary. In the case of a bound secondary pair orbiting the host star (or an escaping pair), we refer to the orbit of the pair around their mutual centre-of-mass as the {\em internal orbit}. We refer to the orbit of the pair around the host star as the {\em external orbit}. 
Two secondary objects with masses $m_1$ and $m_2$, with a relative separation $r_{12}$ and with relative velocity $v_{12}$ are identified as a binary if they satisfy the following criteria: 
\begin{enumerate}
\item they are each other's mutual nearest neighbours,
\item neither of the two objects is the host star,
\item {their total binding energy is negative,
\begin{equation} \label{eq:bindingenergy}
E_b = \frac{1}{2} \mu v_{12}^2 - \frac{Gm_1m_2}{r_{12}} < 0 \ ,
\end{equation}
where $\mu=m_1m_2/(m_1+m_2)$ and $G$ is the gravitational constant, and }
\item the relative separation $r_{12}$ is smaller than their mutual Hill radius $r_H$,
\begin{equation} \label{eq:hill}
r_{12}< r_H \approx \left(\frac{r_1+r_2}{2}\right) \left(\frac{m_1+m_2}{3M}\right)^{1/3} \ , 
\end{equation}
where $M$ is the mass of the host star, and $r_1$ and $r_2$ are the distances of the two companions to the host star, respectively.
\end{enumerate}

At the end of the simulations ($t=10$~Myr) we determine the orbital parameters (the semi-major axis $a$, eccentricity $e$, and inclination $i$) for all remaining single and binary systems orbiting the host star. Orbital inclinations are calculated with respect to the initial orbital plane of each system (which roughly corresponds to the plane of rotation of the host star). The inclinations are expressed in the range $0^\circ \le i \le 180^\circ$, where $i=0^\circ$, $90^\circ$ and $180^\circ$ correspond to purely prograde, polar, and retrograde orbits, respectively. We also calculate the internal orbital parameters of bound and unbound binary systems. During the calculation of the orbital parameters we ignore the presence of other secondary systems orbiting around the host star. A single or binary secondary is considered to be bound to the host star if its binding energy with respect to the host star is negative, and is considered an escaper in all other cases. All escapers are integrated over the entire time span of the simulations, and are almost all found at distant locations ($0.5-50$~pc) from their host star when the simulations have finished at $t=10$~Myr.

A secondary can physically collide with the host star or another secondary. In our model, two bodies with radii $R_1$ and $R_2$ merge when their relative separation $r$ is smaller than the sum of their physical radii: $r<R_1+R_2$. In the case of such an event, both bodies are merged into a single body. The radii of the objects in the simulations are calculated  using the mass-radius relation
\begin{equation} \label{eq:massradiusrelation}
        R(M)=\left\{
        \begin{array}{ll}
        0.1~\rsun & {\rm for~}M<0.1~\msun\\
        (M/\msun)\rsun & {\rm for~}0.1~\msun \le M<1~\msun\\
        \end{array}
        \right.\ ,
\end{equation}
which is a linearisation of the results of \cite{chabrier2009}. All mergers are treated as inelastic. After a merger, the masses of the bodies are combined, and the merger product is assigned the velocity of the centre-of-mass of the two bodies. Subsequently, the radius of the merger product is calculated using Eq.~\ref{eq:massradiusrelation}. 

Since we carry out simulations of isolated systems, dynamical evolution may result in binary or multiple systems with arbitrarily large separations. In reality, extremely wide systems can be destroyed by, for example, the Galactic tide \cite[e.g.,][]{veras2009, jiang2010}, and close encounters with other stars \citep[e.g.][]{tremaine1993}.

%%%%%%%%%%%%%%%%%%%%%%%%%%%%%%%%%%%%%%%%%%%%%%%%%%%%%%%%%%%%%%%%%%%%%%%%%%%%%
%%%%%%%%%%%%%%%%%%%%%%%%%%%%%%%%%%%%%%%%%%%%%%%%%%%%%%%%%%%%%%%%%%%%%%%%%%%%%
%%%%%%%%%%%%%%%%%%%%%%%%%%%%%%%%%%%%%%%%%%%%%%%%%%%%%%%%%%%%%%%%%%%%%%%%%%%%%

\section{Initial conditions} \label{section:initial}

\begin{figure}
  \centering
  \begin{tabular}{cc}
  \includegraphics[width=0.45\textwidth,height=!]{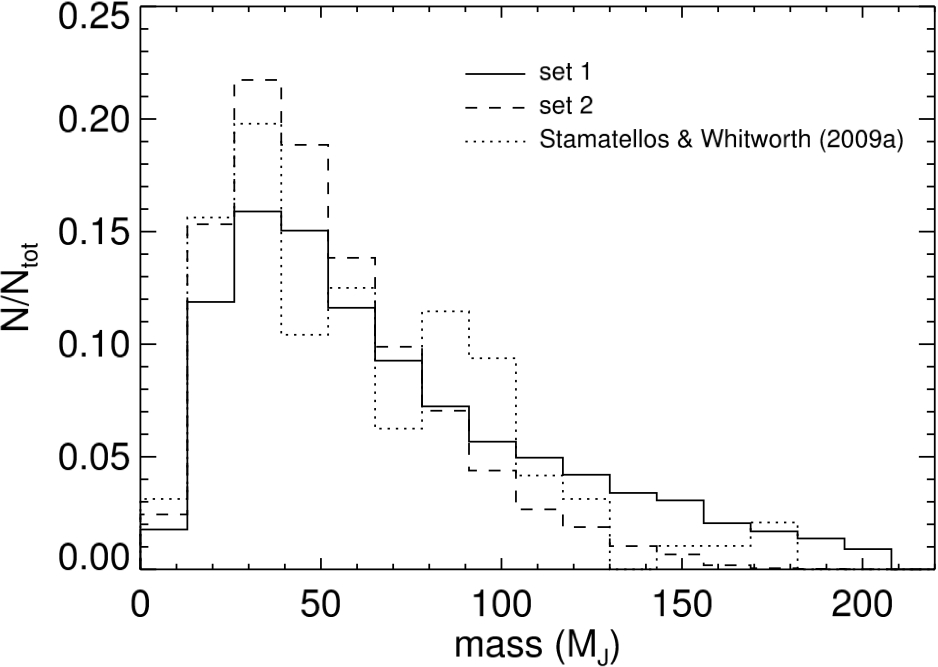} &
  \includegraphics[width=0.45\textwidth,height=!]{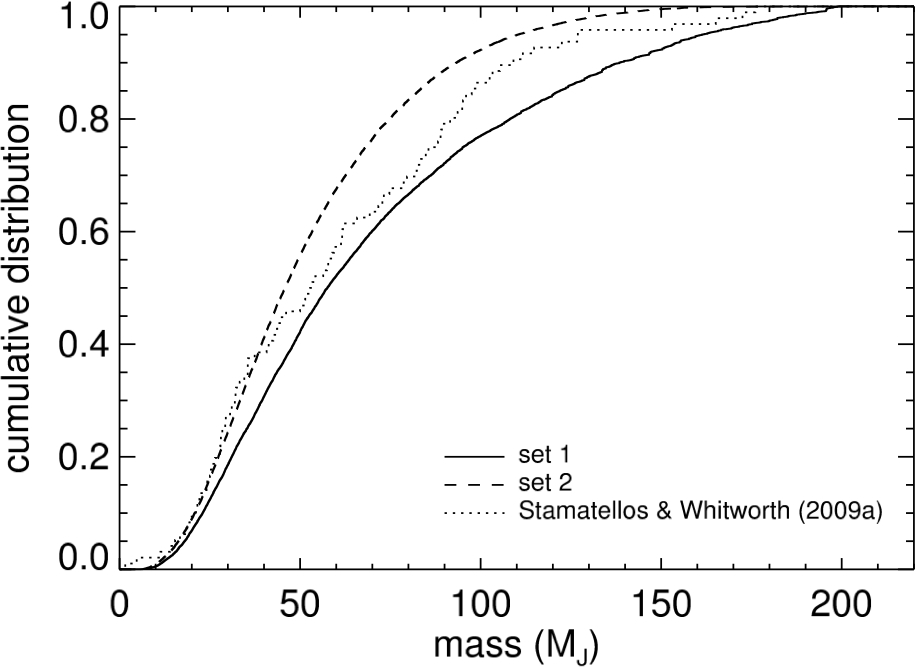} \\
  \end{tabular}
  \caption{The initial mass distribution of the secondary objects for set~1 (solid curves) and set~2 (dashed curves). The results of \protect\cite{stamatellos2009a}, indicated with the dotted curves, are shown for comparison.
  \label{figure:m_initial} }
\end{figure}

\begin{table*}
  \begin{tabular}{lll}
    \hline
    Quantity & Set~1 & Set~2 \\
    \hline
    Mass of host star              & $M=0.7~\msun$                                  & $M=0.7~\msun$ \\
    Notional host star disc mass   & $M_{\rm d}\simeq 0.5~\msun$                    & $M_{\rm d}\simeq 0.2~\msun$ \\
    Number of secondary objects    & $4 \le N \le 11$                               & $3 \le N \le 5$ \\
    Mass of secondary objects      & $1~\mjup \leq m \leq 200~\mjup$                & $1~\mjup \leq m \leq 200~\mjup$ \\
                                   & (at least one object with $m > 80~\mjup$)      & \\ 
    Total mass of secondary objects& $0.48~\msun \leq m_{\rm tot} \leq 0.52~\msun$  & $0.18~\msun \leq m_{\rm tot} \leq 0.22~\msun$ \\
    Semi-major axis                & $50~{\rm AU} < a < 350~{\rm AU}$ ($m<80~\mjup$)& $50~{\rm AU} < a < 250~{\rm AU}$  \\
                                   & $50~{\rm AU} < a < 150~{\rm AU}$ ($m \ge 80~\mjup$)   & \\
    Eccentricity                   & $e=0$                                    & $e=0$ \\
    Inclination                    & $0^\circ < i < 5^\circ$                        & $0^\circ < i < 5^\circ$ \\
    Longitude of the ascending node                 & $0^\circ < \Omega < 360^\circ$& $0^\circ < \Omega < 360^\circ$     \\
    Integration time               & $10$~Myr                                       & $10$~Myr \\
    Number of realisations         & $3000$                                         & $6000$ \\

    \hline
  \end{tabular}
  \caption{Initial conditions for the two sets of simulations described in this paper. The probability distributions of all parameters are described in \S\ref{section:initial}. \label{table:initialtable}}
\end{table*}

We assume that the solar-type stars in our simulations are born with circumstellar discs. These discs subsequently fragment, mainly into brown dwarfs (BDs) but also into low-mass hydrogen-burning stars (LMSs) and into planetary-mass objects (PMOs); see, e.g., \cite{stamatellos2007a}, \cite{stamatellos2009a, stamatellos2009b}, \cite{attwood2009} and \cite{stamatellos2011}. The initial conditions for our simulations are based on the outcomes of the radiation hydrodynamic simulations of \cite{stamatellos2009a}, who used the SPH code DRAGON \citep{goodwin2004a, goodwin2004b} to model the hydrodynamics following the method of \cite{stamatellos2007a} to treat the effects of radiative transfer \citep[see also][]{forgan2009}.

\cite{stamatellos2007a} and \cite{stamatellos2009a} assumed discs with different masses, up to $0.7~\msun$, so that many low-mass objects form in discs to improve the statistical analysis of their results. However, even discs with masses down to $\sim 0.25~\msun$ and radii of order 100~AU can fragment \citep{stamatellos2011}. Such disc masses are comparable to observed disc masses around young protostars \citep[e.g.,][]{tobin2012, favre2014}.
In this paper, we simulate two different sets, which we refer to as set~1 and set~2, respectively. For set~1 we use the outcomes of the simulations of \cite{stamatellos2009a} to construct our initial conditions. For comparison, we also carry out additional simulations with slightly different initial conditions (set~2). The systems in the latter simulations differ from set~1 in terms of secondary masses, semi-major axes, and the number of secondary objects, and are carried out to study the evolution of systems formed from lower-mass circumstellar discs. The latter simulations also allow us to evaluate the robustness of the results on the initial conditions assumed. The initial conditions for both sets of simulations are summarised in Table~\ref{table:initialtable} and described below.

In each system, the host star is assigned a mass of $M=0.7~\msun$ which is almost the same to the host star mass of the simulation of \cite{stamatellos2009a}. As a result of physical collisions, the mass of the host star may grow slightly over the course of time (see Section~\ref{section:collisions}). Each system is randomly assigned $N=4-11$ secondary objects for set~1, and $N=3-5$ secondary objects for set~2.

The mass spectrum of the secondary objects formed in the SPH simulations of \cite{stamatellos2009a} can be approximated by a log-normal distribution and is therefore modelled as such with $\mu=\ln(0.045~\msun)$ and $\sigma=0.65$, with the additional constraints below. To avoid very small or very large value for the mass of the secondary objects, we truncate the above distribution for masses outside the domain $1-200~\mjup$. Both limits correspond to the outcomes of the simulations of \cite{stamatellos2009a}. The lower limit is set by the opacity limit for fragmentation, which is thought to be around $1-3~\mjup$ \citep{whitworth2006}. The upper limit corresponds to a mass limit of an object formed by disc fragmentation, and this may be higher for more massive discs around more massive stars. 
In addition we impose for set~1 that each system contains at least one object with a mass larger than $80~\mjup$ \citep[see][]{stamatellos2009a}, while we do not impose this constraint for set~2. Finally, we impose that the initial total mass of the secondary objects is $0.48-0.52~\msun$ for set~1 and $0.18-0.22~\msun$ for set~2. These values correspond to the mass of host star disc that has fragmented to produce these objects. 
The initial mass distribution of the secondary objects is shown in Figure~\ref{figure:m_initial}.

The simulations of \cite{stamatellos2009a} suggest that the most massive objects that form in the disc, i.e. LMSs, are more likely to form close to the host star. On the other hand, BDs and PMOs may form even at larger distances from the host star. To mimic this behaviour for set~1, we distribute LMSs ($m \ge 80~\mjup$) at distances $50-150$~AU, and BDs and PMOs at distances of $50-350$~AU from the host star. For simplicity we assume flat distributions, i.e. each distance in the above ranges has equal probability to be chosen.  Strong dynamical evolution of the systems at early times (even within one orbital period) quickly broadens any initial semi-major axis distribution. We therefore do not expect a significant difference in the results when other, non-uniform, semi-major axis distributions are used (the systems may decay slightly faster when a more concentrated semi-major axis distribution, such as a Gaussian distribution, is adopted).
The semi-major axis in set~2 for all secondary objects is $50-250$~AU. As all systems consist of a host star of mass $0.7~\msun$ and as the secondary masses are generally much smaller, the orbital periods for all secondaries can be approximated with $P\approx 1.2 (a/{\rm AU})^{3/2}$~yr.

For simplicity we assume the secondary objects initially have circular orbits ($e=0$). We expect that the secondary objects form on roughly circular orbits, since they form from disc gas that moves nearly in Keplerian orbits around the host star. The gas disc may initially provide some damping to the eccentricities of the secondary objects, while it is still massive enough. However, since in general a few objects form in the disc, they will interact with each other and the eccentricities will inevitably rise. For typical secondary mass objects of $0.03~\msun$, the Hill radius is approximately $R_H \approx 0.2 R$ (where $R$ is the distance to the host star), which is relatively large and ensures strong interactions between the secondary objects. These interactions will excite the eccentricities of the orbits of the secondaries.
The initial inclinations $i$ of the orbits of the secondary objects are drawn from a flat distribution in the range $0^\circ-5^\circ$. Due to large number of close encounters, the inclination distribution broadens quickly. Finally, the longitude of the ascending node $\Omega$ is randomised.

We run a large ensemble of simulations in order to obtain a good statistical sample: 3000 runs for set~1 and 6000 runs for set~2. There are more simulations with the set~2 of initial conditions to compensate for the fact that these systems contain fewer secondaries. The total number of secondary objects in the combined sample is $N_{\rm tot,1}=22367$ for set~1 and $N_{\rm tot,2}=23977$ for set~2, respectively. Each simulation is carried out for 10~Myr, beyond which further changes in the remaining system are minor. In other words, by that time, the systems have reached (quasi-) stable configurations.

%%%%%%%%%%%%%%%%%%%%%%%%%%%%%%%%%%%%%%%%%%%%%%%%%%%%%%%%%%%%%%%%%%%%%%%%%
%%%%%%%%%%%%%%%%%%%%%%%%%%%%%%%%%%%%%%%%%%%%%%%%%%%%%%%%%%%%%%%%%%%%%%%%%
%%%%%%%%%%%%%%%%%%%%%%%%%%%%%%%%%%%%%%%%%%%%%%%%%%%%%%%%%%%%%%%%%%%%%%%%%

\section{Dynamical decay of the systems} \label{section:decay}

\begin{figure}
  \centering
  \includegraphics[width=0.45\textwidth,height=!]{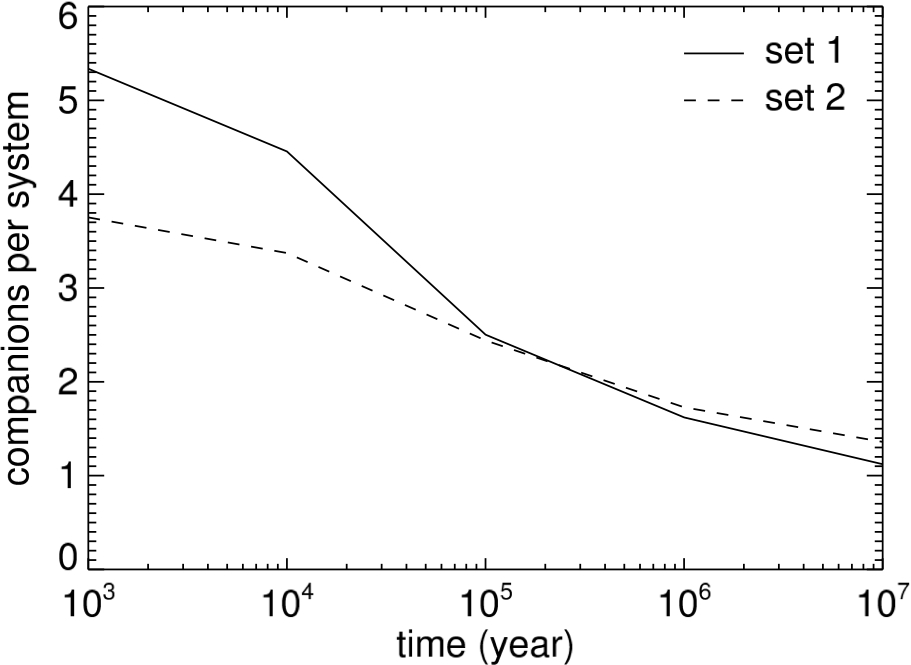}
  \caption{The number of bound secondary objects per system as a function of time for $t \ge 1000$~yr. The solid and dashed curves indicate the averaged results for the ensemble of simulations of set~1 and set~2, respectively. The initial average number of companions per host star, 7.46 and 4.00, respectively, are not shown in this figure.
  \label{figure:number_bound} }
\end{figure}

\begin{table*}
	\centering
  \begin{tabular}{lrr}
    \hline
    Properties at $t=10$~Myr & Set~1  & Set~2 \\
    \hline
    Average number of companions per system   & 1.12                 & 1.36 \\
    \hline
    Single host star                        & $22\,\%$              & $9.8\,\%$ \\
    Host star with 1 secondary              & $48\,\%$              & $46\,\%$ \\
    Host star  with 2 secondaries           & $25\,\%$              & $42\,\%$ \\
    Host star  with 3 secondaries           & $4.3\,\%$             & $2.0\,\%$ \\
    Host star  with 4 secondaries           & $0.33\,\%$            & $0.02\,\%$\\
    Host star  with $\ge 5$ secondaries     & none                  & none\\
    \hline
    Bound singles per system            & $1.1$            & $1.3$\\
    Bound single PMOs per system        & $0.002$          & $0.005$ \\
    Bound single BDs per system         & $0.3$            & $0.8$ \\
    Bound single LMSs per system        & $0.7$            & $0.5$ \\
    \hline
    Escaped singles per system          & $5.5$               & $2.4$\\
    Escaped single PMOs per system      & $0.12$             & $0.09$ \\
    Escaped single BDs per system       & $4.0$               & $2.2$ \\
    Escaped single LMSs per system      & $1.3$               & $0.13$ \\
    \hline
    Escape fraction of PMOs        & $98\,\%$       & $95\,\%$ \\
    Escape fraction of BDs         & $93\,\%$       & $72\,\%$ \\
    Escape fraction of LMSs        & $64\,\%$       & $21\,\%$ \\
	\hline
    Bound binaries per system                        & $0.033$               & $0.018$ \\
    Bound PMO-PMO binaries per system                & none                  & none\\
    Bound PMO-BD binaries per system                 & none                  & $<0.001$\\
    Bound PMO-LMS binaries per system                & none                  & none\\
    Bound BD-BD binaries per system                  & $0.005$               & $0.010$\\
    Bound BD-LMS binaries per system                 & $0.013$               & $0.008$\\
    Bound LMS-LMS binaries per system                & $0.014$               & $<0.001$\\
    \hline
    Escaped binaries per system                      & $0.182$             & $0.026$ \\
    Escaped PMO-PMO binaries per system              & none                & none\\
    Escaped PMO-BD binaries per system               & $0.004$             & $0.002$\\
    Escaped PMO-LMS binaries per system              & $0.001$             & $<0.001$\\
    Escaped BD-BD binaries per system                & $0.087$             & $0.020$\\
    Escaped BD-LMS binaries per system               & $0.064$             & $0.004$\\
    Escaped LMS-LMS binaries per system              & $0.026$             & none\\
    \hline
    Binary fraction among all secondaries            & $3.2\,\% $       & $1.2\,\% $ \\
    Binary fraction among bound secondaries          & $3.0\,\% $       & $1.3\,\% $ \\
    Binary fraction among escaped secondaries        & $3.2\,\% $       & $1.1\,\% $ \\
    \hline
    Fraction of PMOs part of a binary system      & $3.3\,\%$       & $2.3\,\%$\\
    Fraction of BDs part of a binary system       & $5.8\,\%$       & $2.4\,\%$\\
    Fraction of LMSs part of a binary system      & $7.1\,\%$       & $1.8\,\%$\\
    \hline
    Collisions with the host star per system          &  0.43  & 0.18 \\
    Collisions between secondaries per system         &  0.08  & 0.04 \\
    \hline
  \end{tabular}
  \caption{Statistical properties of the systems at $t=10$~Myr. \label{table:leftover} }
\end{table*}

\begin{figure*}
  \centering
  \begin{tabular}{cc}
  \includegraphics[width=0.45\textwidth,height=!]{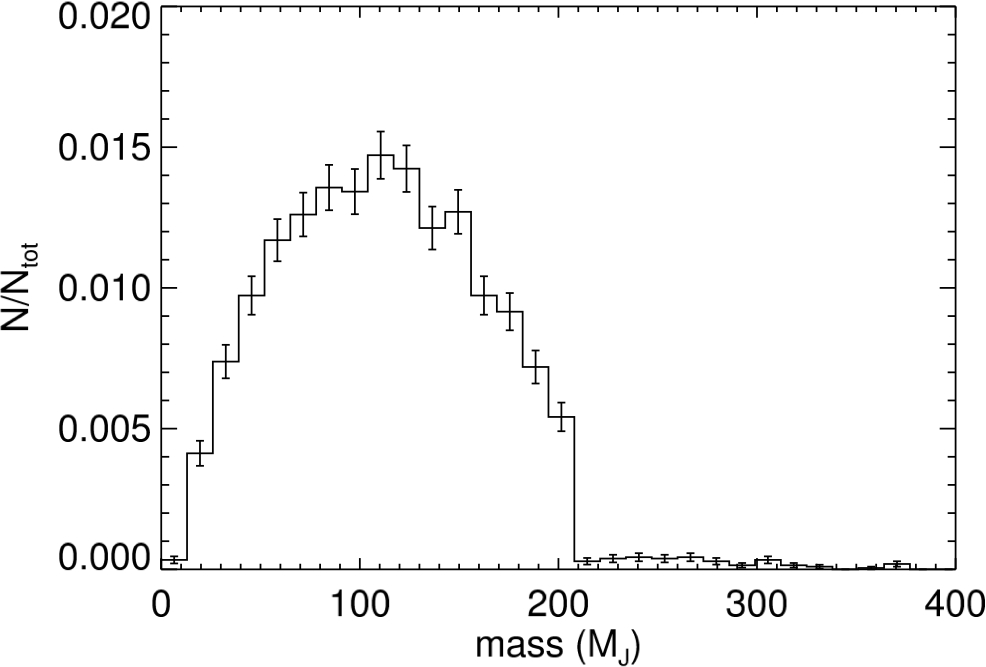} &
  \includegraphics[width=0.45\textwidth,height=!]{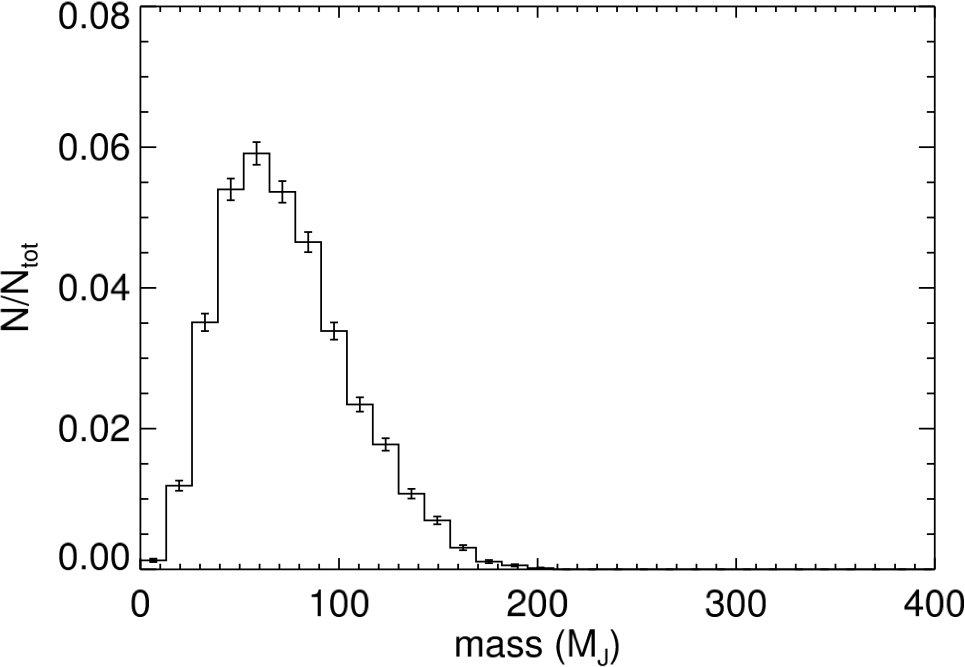} \\
  \end{tabular}
  \caption{The mass distribution of bound secondary objects at $t=10$~Myr for set~1 ({\em left}) and set~2 ({\em right}), with Poissonian errors indicated. The distributions are normalised to the total number of companions $N_{\rm tot}$. All objects with masses larger than $200~\mjup$ are the results of physical collisions between secondaries. 
  \label{figure:m_bound} }
\end{figure*}

All modelled systems are initially unstable and evolve quickly. The decrease in the number of bound secondaries is primarily caused by escape, and for a minimal amount by physical collisions. Figure~\ref{figure:number_bound} shows for both sets of simulations how the number of bound secondary objects per system evolves over time. The final configurations of the systems are summarised in Table~\ref{table:leftover}. The average initial number of secondaries per system is 7.46 for set~1 and 4.00 for set~2. Approximately half of the secondaries escape or collide within 20\,000 yrs for set~1 and within 400\,000 yrs for set~2. Beyond $t=1$~Myr, the number of bound secondaries continues to decrease. This decrease slows down as time passes, indicating that little dynamics of interest occurs beyond $t=10$~Myr. Although the systems in set~1 initially have almost twice as many secondaries as those in set~2, the decay rate for set~1 is stronger and results in, on average, a smaller number of companions per system ($\approx 1.1$) in the final configuration at $t=10$~Myr than for set~2 ($\approx 1.4$).

Ejections from non-hierarchical, equal-mass multiple systems typically occur beyond 100 crossing times \citep{anosova1986}. Although our systems are neither non-hierarchical nor equal-mass, they rapidly decay, and it is insightful to look at our results in the context of the work of \cite{anosova1986}. The rate at which secondaries escape from their host star is closely related to the rate at which close encounters occur. The encounter rate is inversely proportional to the typical orbital period of the companions, and is larger when more companions are present. Following a close encounter that results in the escape of a secondary, the crossing time increases, as there is now one less companion in the system (although the orbital periods may be somewhat shorter). As close encounters occur less frequently in the remaining system, it takes longer for the next secondary to escape. This ultimately results in an almost flat curve in Figure~\ref{figure:number_bound} at large times. At $t=10$~Myr, the majority of the host stars are single or have one companion left. In the cases where more than one companion remains bound, the remaining companions have formed bound binary companions that orbit the host star, or reside on widely separated orbits.
In the latter multiple systems, the companions are separated by at least several, and sometimes many mutual Hill radii (Eq.~\ref{eq:hill}, where $r_1$ is the apastron distance of the inner companion and $r_2$ the periastron distance of the outer companion). For typical companion masses in our systems, this means that the semi-major axis of the outer companion is at least several times larger than that of the inner one.

Due to strong gravitational interactions between secondaries, the orbital configuration of the remaining bound secondaries changes much during the first 10~Myr (see Section~\ref{section:elements}). At $t=10$~Myr, approximately 22\% and 10\% of the systems in set~1 and set~2 have no secondary objects left, while roughly 48\% and 46\% of the systems in set~1 and set~2 respectively still have one companion left. For the systems with two secondary objects left (roughly 25\% and 42\%, respectively), the orbits of the two secondary objects are either well separated, or the two secondary objects have paired up in a circumstellar binary, and have a stable triple configuration (see the discussion in Section~\ref{section:binarity}). Only a small fraction of the systems has three or even more secondary objects left at 10~Myr. The orbits of these secondaries are again well separated, or several of the secondaries have paired up into binary companions. Therefore even if $4-11$ LMSs, BDs, and PMOs form in each disc as \cite{stamatellos2009a} suggest, only few of them will remain bound to the host stars at 10~Myr. This is consistent with observations that no stars with more than a few companion BDs have been discovered so far \citep{mason2015}.

The mass distributions of the bound secondaries at $t=10$~Myr are shown in Figure~\ref{figure:m_bound}. Scattering of secondary objects results in the preferred ejection of lower-mass secondary objects (cf. the initial mass distributions in Figure~\ref{figure:m_initial}). Over a thousand secondary objects have collisions with the host star, and over two hundred merger events following collisions between secondary objects occur for each set of simulations. As a result of physical collisions some host stars and secondaries obtain larger masses. For set~1 in particular, some collisions between secondaries result in the formation of secondaries with masses larger than the initial secondary upper mass limit of $200~\mjup$. The physical collisions are discussed in further detail in Section~\ref{section:collisions}.

The majority of the secondaries escape from their host star within 10~Myr. Only a small fraction ($2\%$ for set~1 and $5\%$ for set~2) of the PMOs remain bound to the host star in both simulations. This is expected as these are the lowest mass objects in the system and are therefore more likely to be the ones that are ejected during three-body encounters. This casts doubt on whether wide-orbit giants \citep[e.g.,][]{marois2008} may form by disc fragmentation; if they do, then they have to be the largest mass object in the disc to avoid ejection. It is still uncertain whether such low-mass objects may form in lower-mass discs and remain bound to the host star, although it seems more likely for PMOs to remain bound in set~2 (5\%, against 2\% in set~1). On the other hand, free-floating giant planets, which may be quite common \citep[e.g.,][]{sumi2011}, may form in discs and get ejected into the field; this happens routinely in our simulations. 
Low-mass binary companions have been discovered orbiting solar-type stars \citep{faherty2010, faherty2011, burgasser2012}, which is qualitatively the outcome of our simulations. However, in our models the vast majority of the BDs escape as singles, while a much smaller fraction escapes as part of a binary companion, remains bound to the hosts star as a binary companion, or remains bound to the host star as a single BD. The properties of the escaping single and binary secondaries are further discussed in Section~\ref{section:escapers}.

Observations of young stellar populations indicate that the binary fraction decreases with primary star mass \citep[see, e.g., the review by][]{duchene2013a}, ranging from almost 100\% for early-type stars \citep[e.g.,][]{shatsky2002, kouwenhoven2005, kouwenhoven2007, kobulnicky2014} to $\sim 60\%$ for solar-type stars, to $30-40\%$ for M0-M6 dwarfs \citep{delfosse2004, reid1997, fischer1992}. All host stars in our models are K-type dwarfs, with slightly lower mass than the sample of G-type dwarfs in the multiplicity studies of \cite{duquennoy1991} and \cite{raghavan2010}. However, a comparison with such observational results should be carried out with caution, given the limitations in our models, observational biases in the binarity surveys, and the possibility of other formation mechanisms of companions to K-type dwarfs.

%%%%%%%%%%%%%%%%%%%%%%%%%%%%%%%%%%%%%%%%%%%%%%%%%%%%%%%%%%%%%%%%%%%%%%%%%
%%%%%%%%%%%%%%%%%%%%%%%%%%%%%%%%%%%%%%%%%%%%%%%%%%%%%%%%%%%%%%%%%%%%%%%%%
%%%%%%%%%%%%%%%%%%%%%%%%%%%%%%%%%%%%%%%%%%%%%%%%%%%%%%%%%%%%%%%%%%%%%%%%%

\section{Physical collisions, orbital elements evolution, escape, and binarity} \label{section:other}

\subsection{Physical collisions} \label{section:collisions}

\begin{figure}
  \centering
  \includegraphics[width=0.45\textwidth,height=!]{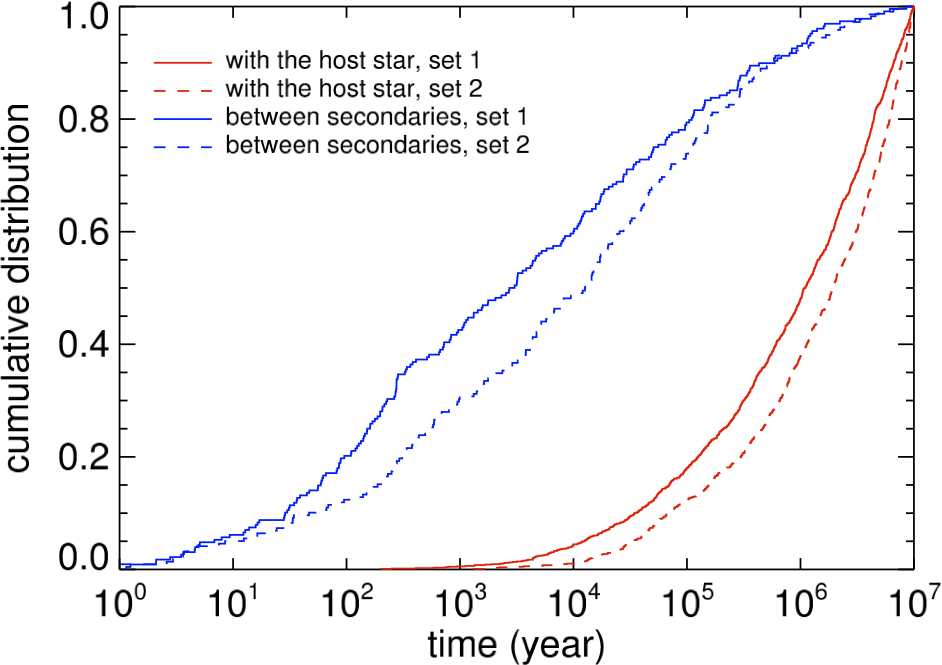}
  \caption{The cumulative distributions of the number of collisions with the host star (red) and between secondaries (blue) as a function of time, for set~1 (solid curves) and set~2 (dashed curves).
  \label{figure:hit_collision} }
\end{figure}

As predicted by \cite{rawi2012}, gravitational scattering in BD-hosting star systems often results in secondary objects colliding with the host star or with each other. In our simulations, the total number of collisions with the host star is 1276 (set~1) and 1055 (set~2), and the total number of collisions between secondaries is 228 and 218 for set~1 and set~2, respectively. Figure~\ref{figure:hit_collision} shows the cumulative distributions of collisions with the host star (red curves) and between secondaries (blue curves) as a function of time.

Despite the different setup of both datasets, the distributions over time are rather similar. Mergers between secondaries mostly occur at early times when few secondaries have escaped (cf. Figure~\ref{figure:number_bound}). Half of the collisions between secondaries occur within about 2~kyr for set~1, which corresponds to merely $1-2$ orbital periods. For set~2, the collisions between secondaries take place at later times, due to the smaller initial companions: half of the collisions between secondaries occur within 10~kyr, which corresponds to roughly 8~orbital periods of the companions. Collisions with the host star mostly occur at later times; roughly half of these occur within about 1~Myr for both datasets, which corresponds to a roughly a thousand initial orbital periods. Collisions with the host star therefore can be attributed to the evolution of marginally stable orbits, either to strong scattering events, or due to secular processes such as the Kozai mechanism.

Each system experiences on average 0.43 (set~1) and 0.18 (set~2) collisions with the host star. As the median companion mass is of order $50~\mjup$, the host stars typically experience an increase in mass of 3\% (set~1) and 1\% (set~2). As the companions are formed from the same circumstellar disc as the host star, these collisions are unlikely to result in observable metallicity changes in the host star atmosphere.
The respective number of mergers between secondary objects are substantially rarer: 0.08 and 0.04 per system. Since the average number of secondary objects per system is smaller for set~2, we expect both the number of collisions with the host star and the number of collisions between secondaries to be smaller for set~2. The fraction of secondary objects involved in a collision is 7.7\% for set~1 and 6.2\% for set~2. These mergers are responsible for the origin of secondaries above our initial mass limit of $200~\mjup$ in Figure~\ref{figure:m_bound}, and interestingly, for the formation of several hydrogen-burning secondaries from BDs.

%%%%%%%%%%%%%%%%%%%%%%%%%%%%%%%%%%%%%%%%%%%%%%%%%%%%%%%%%%%%%%%%%%%%%%%%%
%%%%%%%%%%%%%%%%%%%%%%%%%%%%%%%%%%%%%%%%%%%%%%%%%%%%%%%%%%%%%%%%%%%%%%%%%
%%%%%%%%%%%%%%%%%%%%%%%%%%%%%%%%%%%%%%%%%%%%%%%%%%%%%%%%%%%%%%%%%%%%%%%%%

\subsection{Evolution of the orbital elements of bound secondaries} \label{section:elements}

\begin{figure*}
  \centering
  \begin{tabular}{cc}
  \includegraphics[width=0.45\textwidth,height=!]{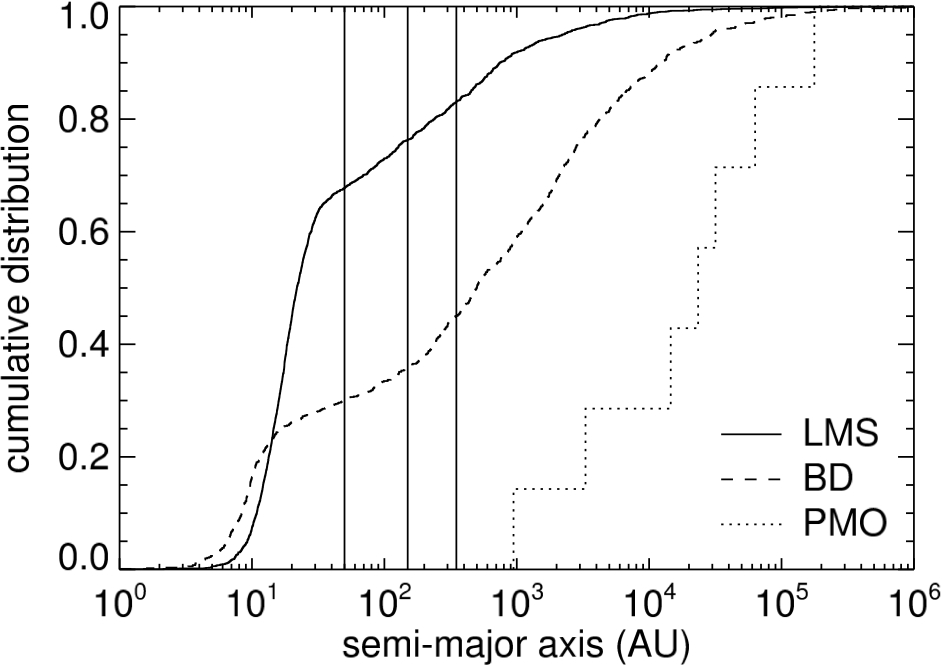} &
  \includegraphics[width=0.45\textwidth,height=!]{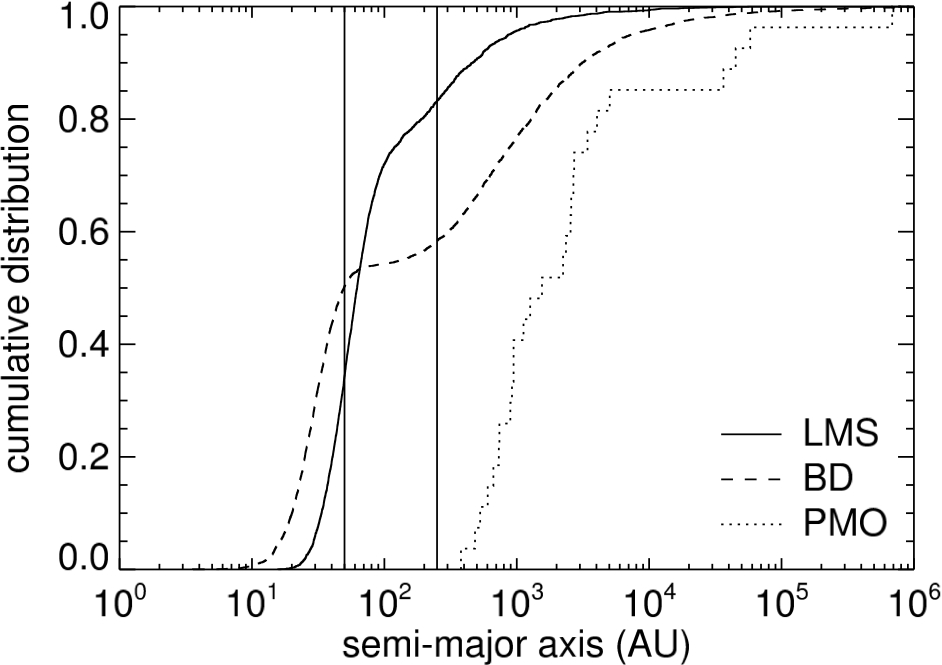} \\
  \includegraphics[width=0.45\textwidth,height=!]{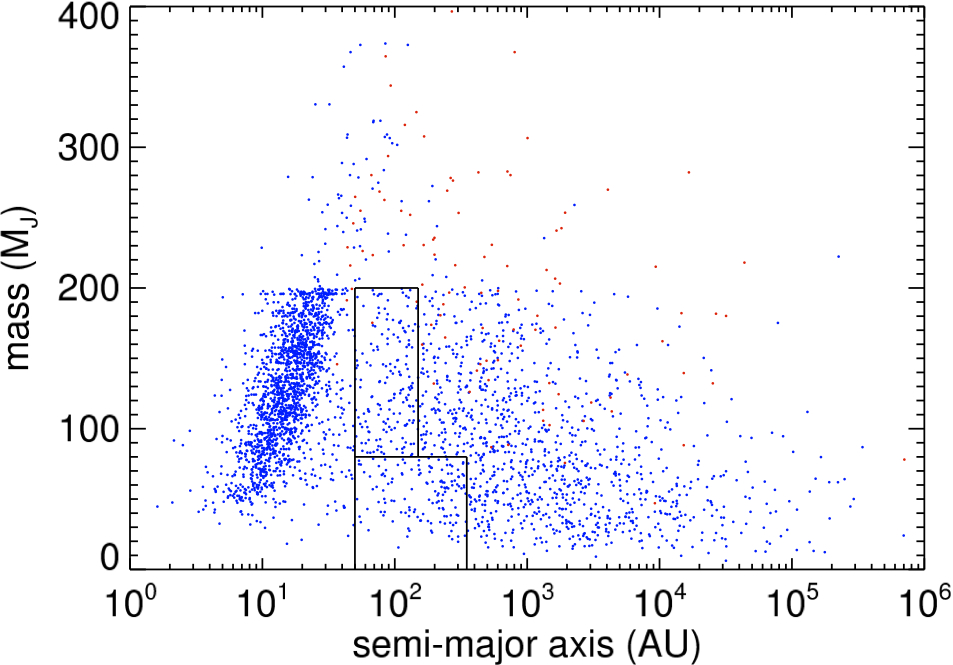} &
  \includegraphics[width=0.45\textwidth,height=!]{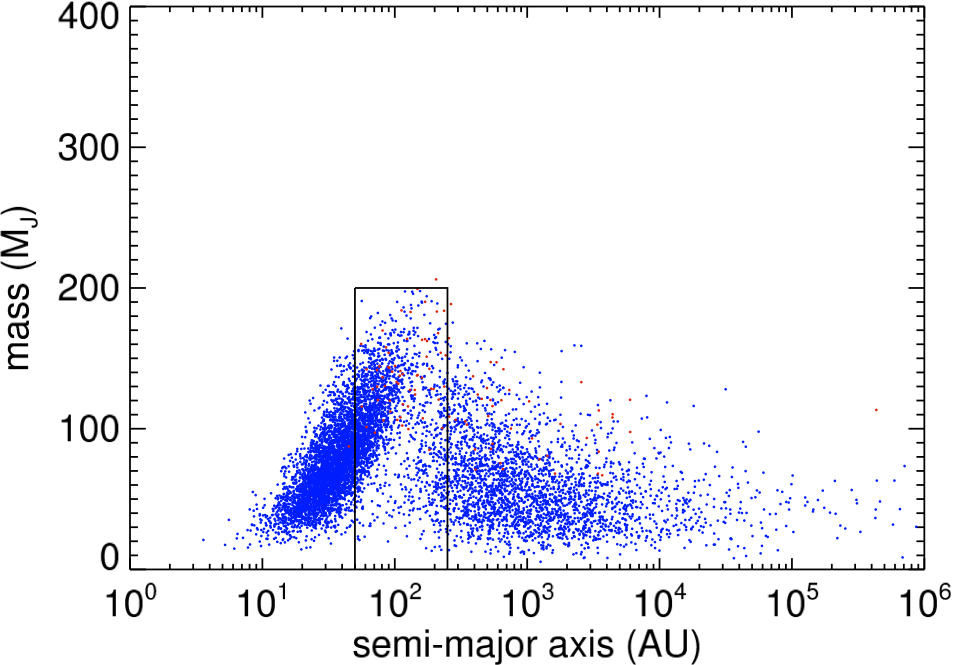} \\
  \end{tabular}
  \caption{Semi-major axis distributions of bound secondaries at $t=10$~Myr for set~1 ({\em left}) and set~2 ({\em right}). {\em Top:} the cumulative distributions of the semi-major axis for LMSs (solid curves), BDs (dashed curves), and PMOs (dotted curves). {\em Bottom:} mass versus semi-major axis. Blue and red dots represent single and binary secondary objects, respectively. Solid lines indicate the boundaries of the initial conditions (see Table~\ref{table:initialtable}).
  \label{figure:a_single} }
\end{figure*}

\begin{figure}
  \centering
  \begin{tabular}{cc}
  \includegraphics[width=0.45\textwidth,height=!]{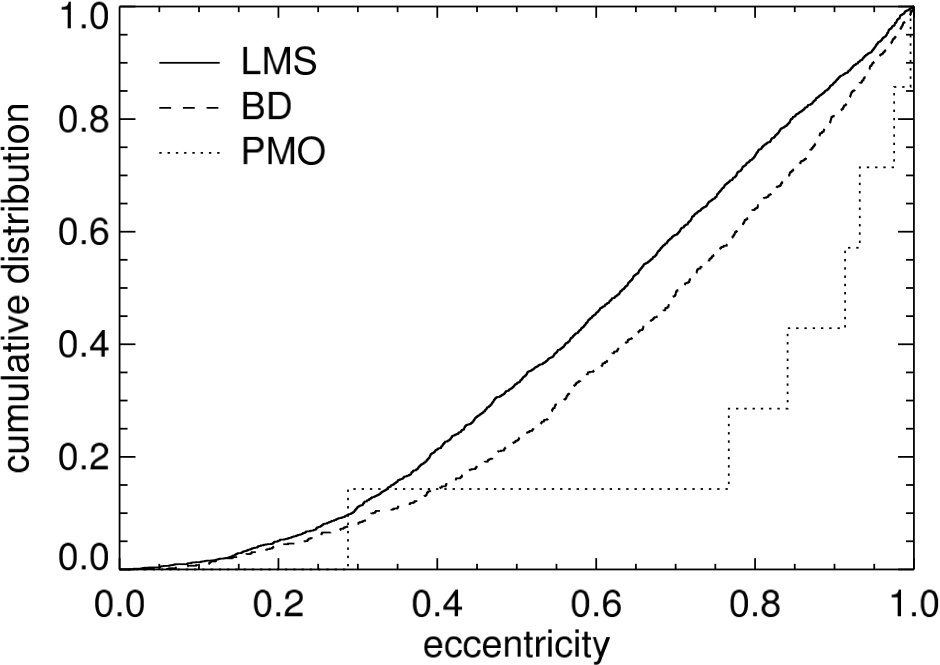} &
  \includegraphics[width=0.45\textwidth,height=!]{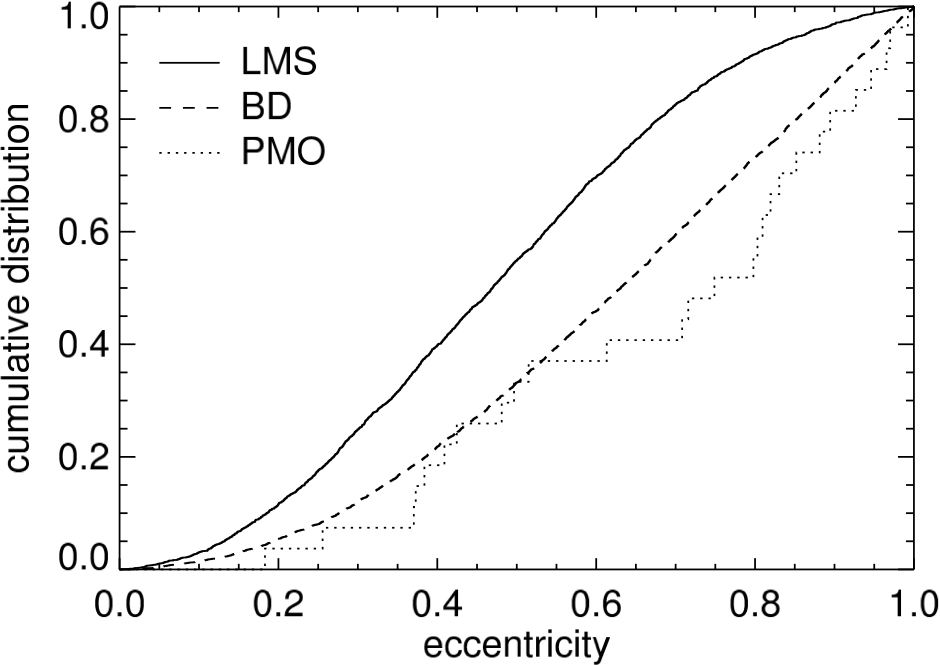} \\
  \includegraphics[width=0.45\textwidth,height=!]{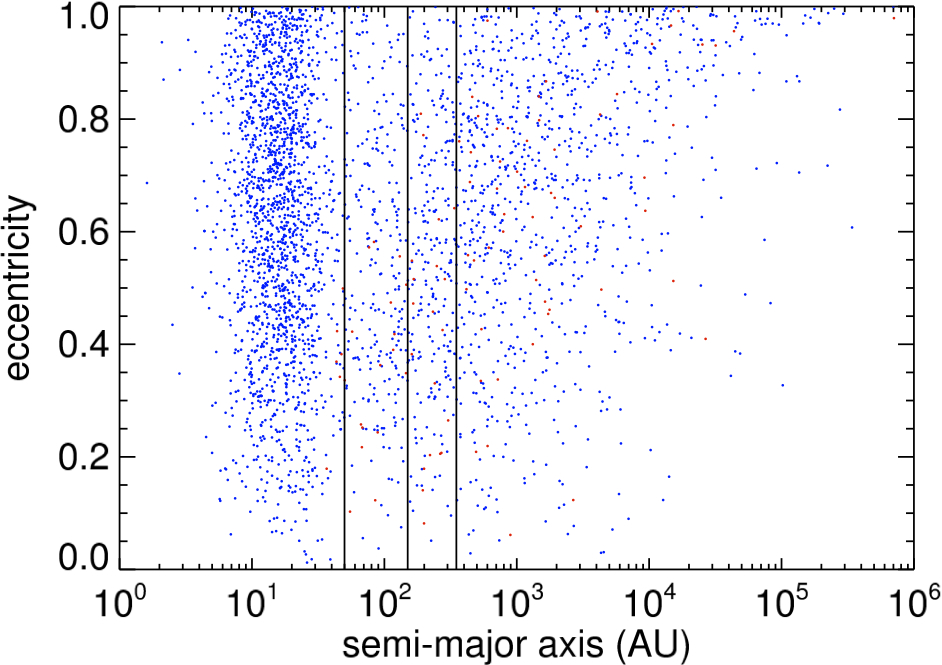} &
  \includegraphics[width=0.45\textwidth,height=!]{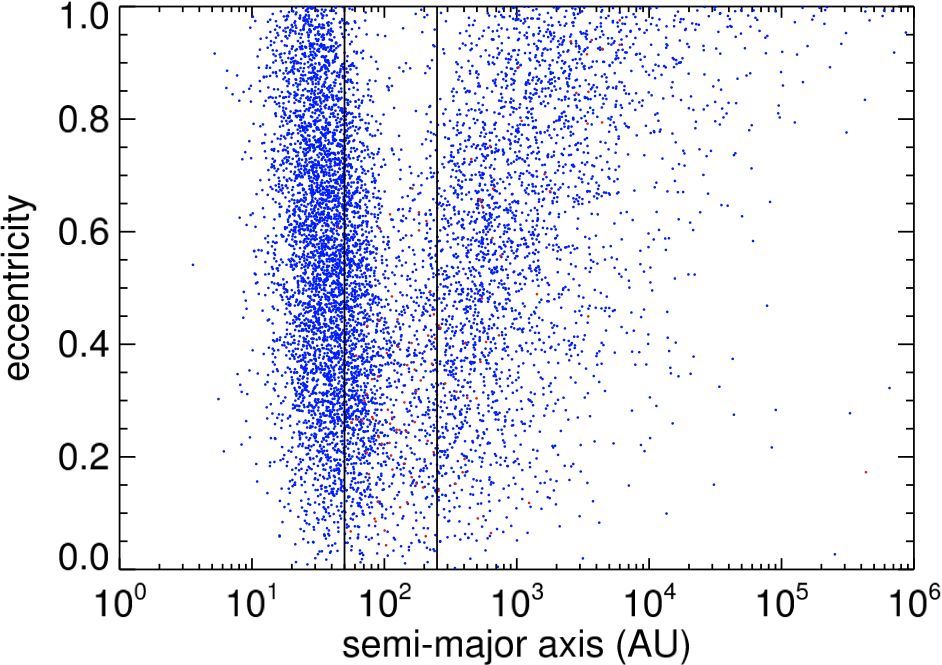} \\
  \includegraphics[width=0.45\textwidth,height=!]{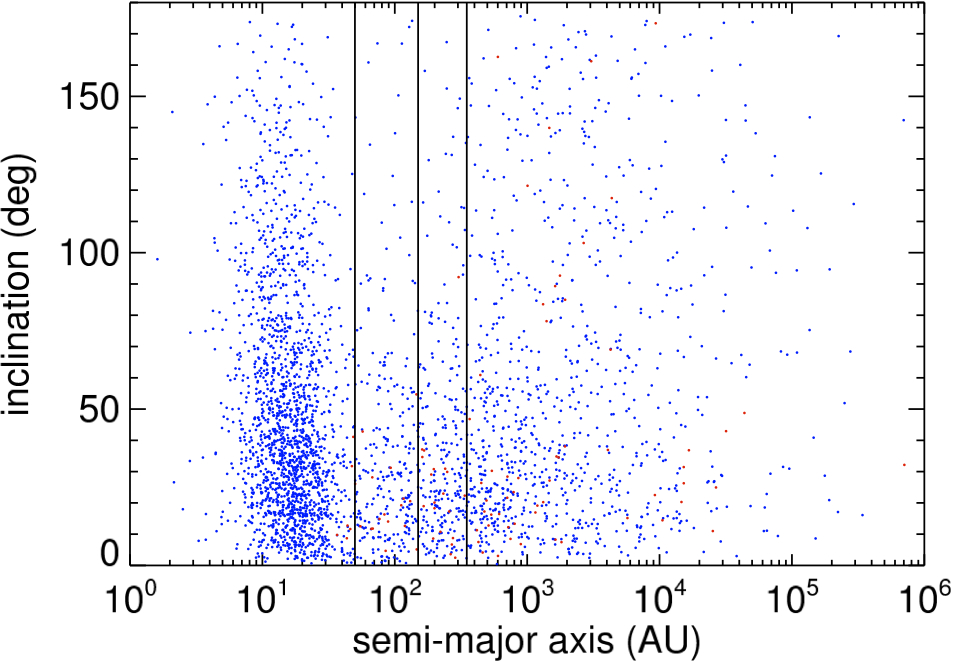} &
  \includegraphics[width=0.45\textwidth,height=!]{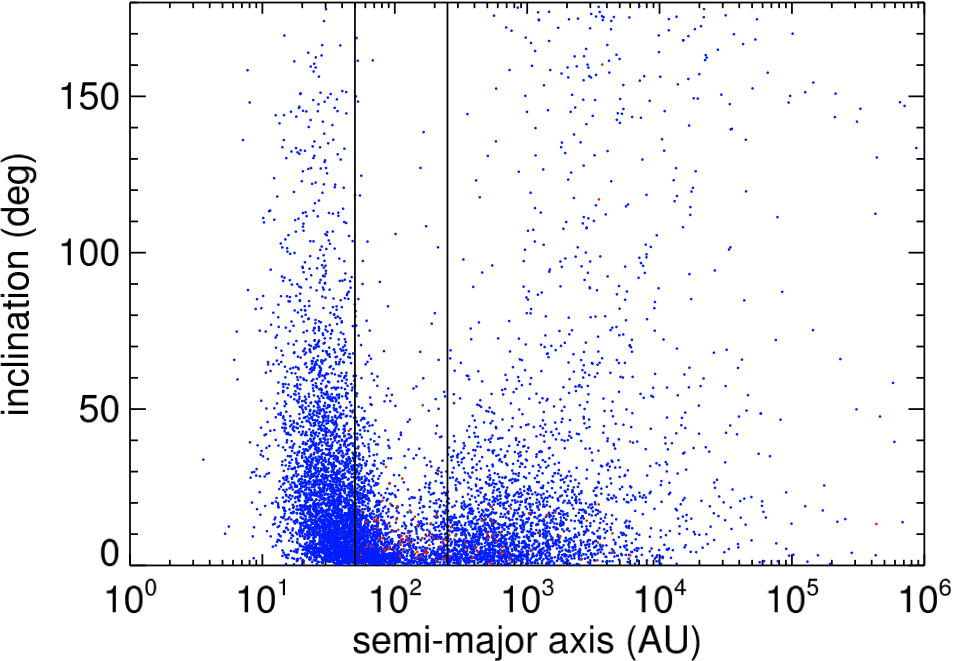} \\
  \end{tabular}
  \caption{Orbital elements of the bound secondaries at $t=10$~Myr for set~1 ({\em left}) and set~2 ({\em right}). {\em Top}: cumulative eccentricity distributions of the LMSs (solid curves), BDs (dashed curves) and PMOs (dotted curves). {\em Middle:} eccentricity versus semi-major axis. {\em Bottom:} external inclinations versus semi-major axis. Blue and red dots represent single and binary secondary objects, respectively. The solid vertical lines represent the boundaries of the initial conditions. 
  \label{figure:ae} }
\end{figure}

\begin{figure}
  \centering
  \includegraphics[width=0.45\textwidth,height=!]{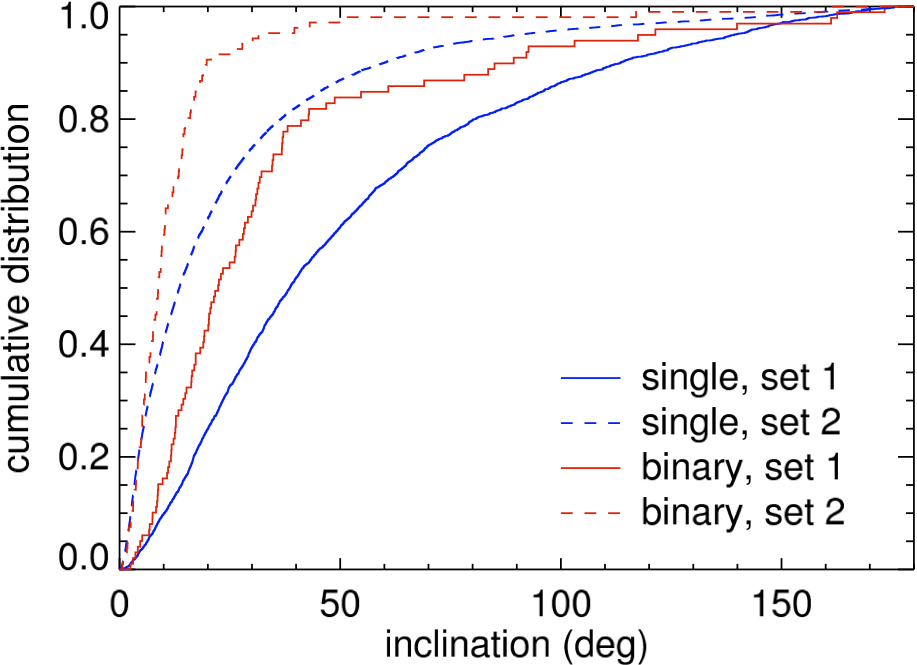}
  \caption{The cumulative inclination distributions for the bound secondaries at $t=10$~Myr for set~1 (solid curves) and set~2 (dashed curves). Blue and red curves represent single and binary secondary objects, respectively. In the case of binary companions, the external inclination is shown, i.e., that of the orbit of the binary pair around the host star. 
  \label{figure:i_out} }
\end{figure}

\begin{figure*}
  \centering
  \begin{tabular}{cc}
  \includegraphics[width=0.45\textwidth,height=!]{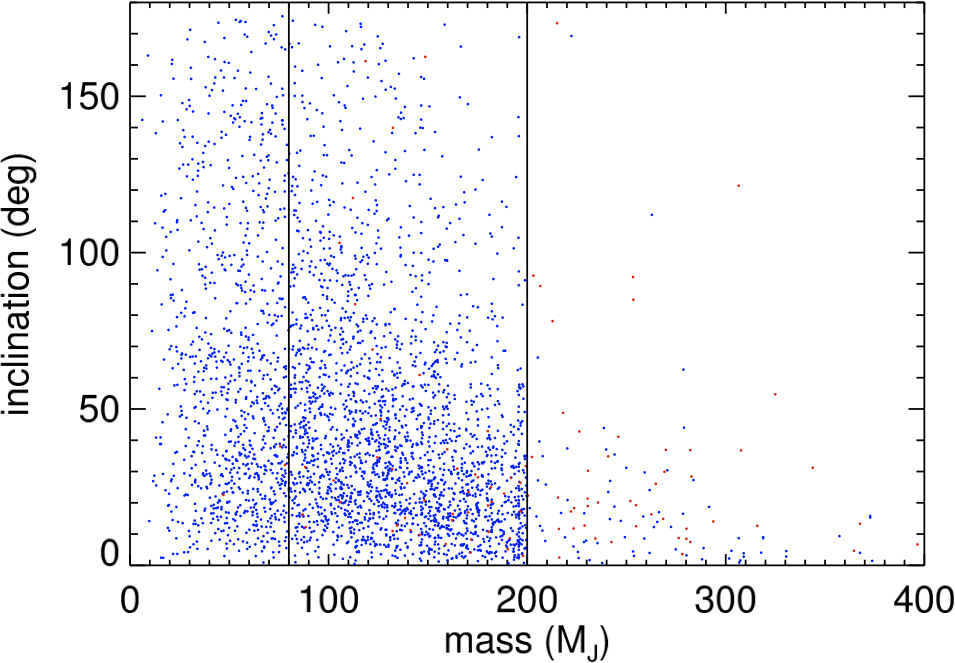} &
  \includegraphics[width=0.45\textwidth,height=!]{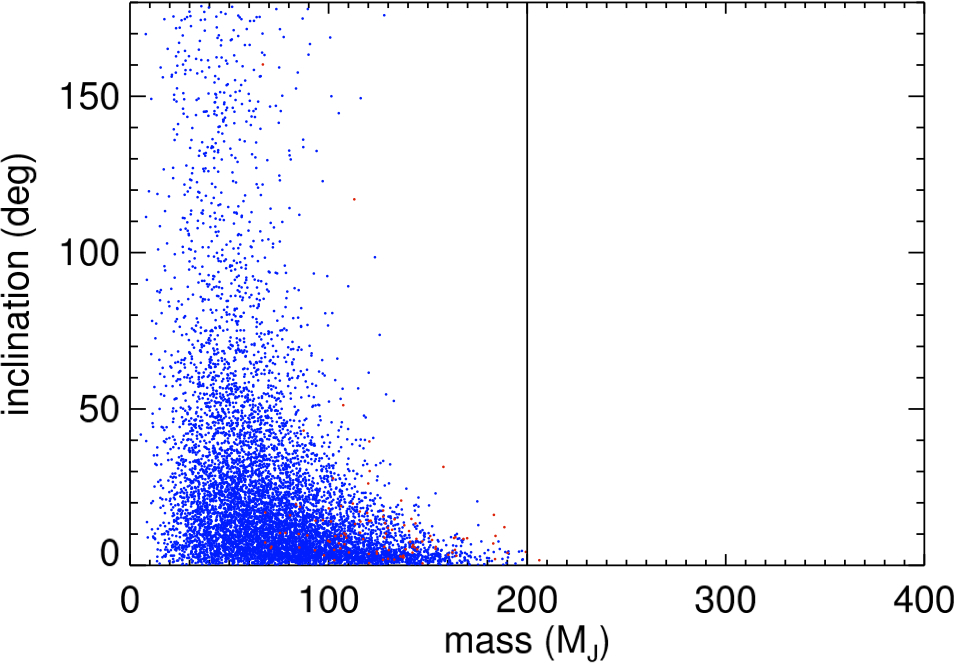} \\
  \end{tabular}
  \caption{External inclination versus mass at $t=10$~Myr for set~1 ({\em left}) and set~2 ({\em right}). Blue and red dots represent single and binary secondary objects, respectively. The solid lines represent the boundaries of the initial conditions.
  \label{figure:mi} }
\end{figure*}

\begin{figure*}
  \centering
  \begin{tabular}{cc}
  \includegraphics[width=0.45\textwidth,height=!]{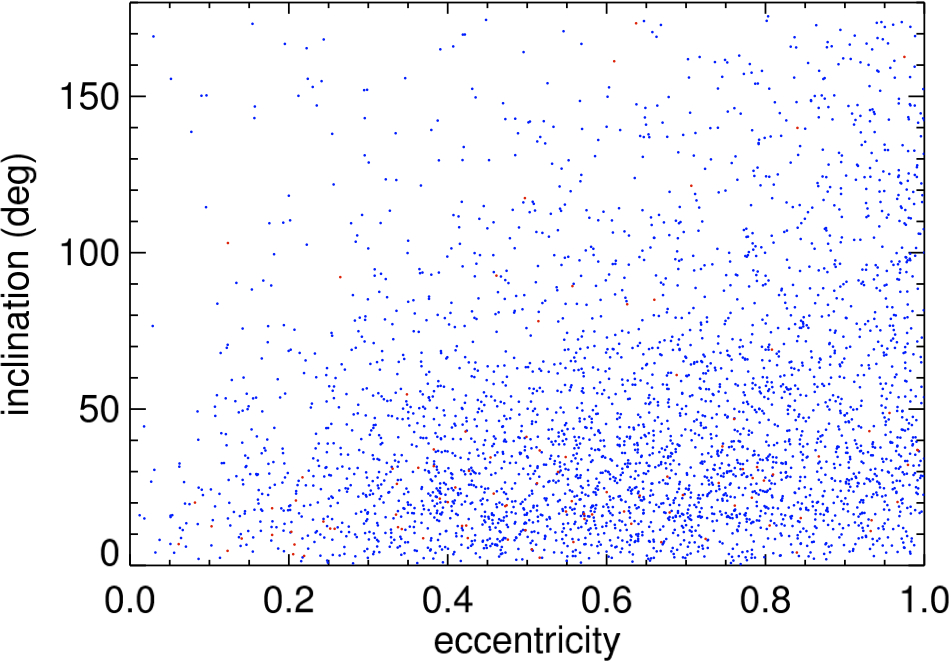} &
  \includegraphics[width=0.45\textwidth,height=!]{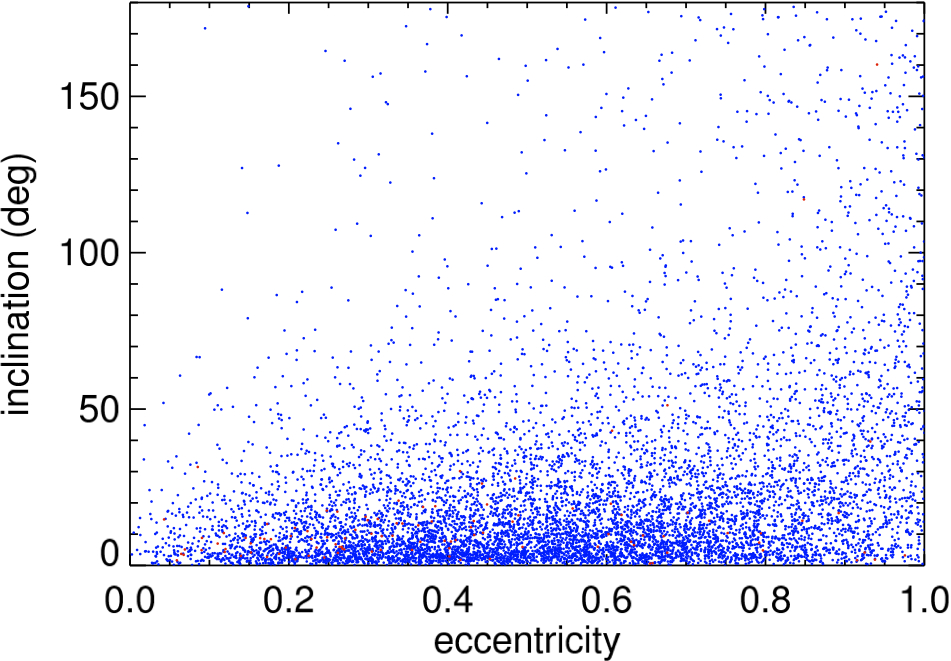} \\
  \end{tabular}
  \caption{External inclination versus eccentricity for the bound secondaries at $t=10$~Myr for set~1 ({\em left}) and set~2 ({\em right}). Blue and red dots represent single and binary secondary objects, respectively.
  \label{figure:ei} }
\end{figure*}

\begin{table*}
	\centering
  \begin{tabular}{llrr}
    \hline
    \multicolumn{2}{l}{Close secondaries at $t=10$~Myr} & set~1 & set~2 \\
  \hline
    Bound  PMOs per system      & $a \le 10$~AU    & none            & none\\
                                      & $a \le 20$~AU    & none            & none \\
                                      & $a \le 50$~AU    & none            & none \\
                                      \hline
    Bound  BDs per system       & $a \le 10$~AU    & $0.05$        & $0.0055$ \\
                                      & $a \le 20$~AU    & $0.08$        & $0.08$ \\
                                      & $a \le 50$~AU    & $0.09$        & $0.41$ \\
                                      \hline
    Bound  LMSs per system      & $a \le 10$~AU    & $0.06$        & none \\
                                      & $a \le 20$~AU    & $0.34$        & $0.0015$ \\
                                      & $a \le 50$~AU    & $0.51$        & $0.17$ \\
    \hline
  \end{tabular}
  \caption{The number of close bound PMOs, BDs and LMSs per system at $t=10$~Myr.
  \label{table:small_a} }
\end{table*}

The semi-major axis distributions at $t=10$~Myr for the PMOs, BDs and LMSs are shown in Figure~\ref{figure:a_single}. Results are shown for both single and binary companions. Each  panel clearly indicates that most secondaries migrate inwards or outwards, and that the rate and direction of this migration strongly depends on mass. This is most clearly seen for set~2 (right-hand panels), where secondaries of all masses initially share an identical semi-major axis distribution. Our model therefore predicts that companions formed through disc fragmentation can be found at semi-major axes smaller than or larger than that at which they had originally formed. 

As LMSs are the heaviest among the secondaries, their separation distribution is least affected by scattering events. Due to the relatively small number of secondaries in set~2, only $\sim 35\%$ of the LMSs migrate inwards of 50~AU, while $\sim 15\%$ migrate outwards of 250~AU. Scattering plays a more prominent role in set~1, where $\sim 65\%$ and $\sim 25\%$ of the LMSs move to radii within and outside the limits 50~AU and 150~AU of the initial conditions, respectively. PMOs, on the other hand, are more vulnerable to being scattered to larger distances, and none of the PMOs in both sets of simulations is able to remain within the initial separation range. The BDs, which are most abundant among the secondaries, show an intermediate behaviour. The widest-orbit secondaries have semi-major axes beyond $10^5$~AU ($0.48$~pc), and are likely to be disrupted due to interactions with nearby stars or the Galactic tidal field \citep{jiang2010}.

The Galactic field hosts a population of wide ($\ga 500$~AU), low-mass binary companions \citep{gizis2001, dhital2010, rodriguez2012}. Such systems can be formed through the decay of disc fragmented systems, as described in this study. They can also originate from capture in star clusters \citep[e.g.,][]{kouwenhoven2010, moeckel2011, parker2012, perets2012}, through the decay of small-$N$ systems \citep[e.g.,][]{clarke1996, sterzik2003}, through turbulent core fragmentation \citep[][]{jumper2013} or through the decay of embedded triple systems \cite[][]{reipurth2012}. Although the model described in the current study is unlikely the dominant mechanism for the origin of very wide low-mass companions to stars, it may contribute to some extent, and observed orbital parameter distributions may be used to distinguish between the different possible formation mechanisms.

The relation between semi-major axis and mass at $t=10$~Myr is shown in the bottom panels of Figure~\ref{figure:a_single}. During a close encounter between two secondary objects, orbital energy is exchanged, but the total energy of the system is conserved. As a result, one of the two secondaries moves closer to the host star, while the other obtains a larger semi-major axis or completely escapes from the system. The final semi-major axis distribution of the bound secondaries is therefore expected to be bi-modal, which is clearly seen in the bottom panels of Figure~\ref{figure:a_single}. The critical distance that separates these two populations occurs at $a\approx 50$~AU for set~1 and $a\approx 150$~AU for set~2. Most bound secondaries in set~1 have migrated to orbits with semi-major axes below the initial lower limit of the semi-major axis distribution (50~AU) as a consequence of outward scattering of (mostly) lower-mass secondaries. The most massive secondaries are more inert to scattering and therefore tend to remain closer to their original locations, while the lower-mass secondaries obtain a wide range of orbital periods or get ejected. This trend is also seen for the binary secondaries that form during dynamical interactions. All secondaries with masses larger than $200~\mjup$ (the initial upper mass limit) are either binary pairs or merger products.

In Table~\ref{table:small_a} we list the number of close LMS, BD, and PMO companions per system at $t=10$~Myr. As none of the secondaries have separations equal to or less than 50~AU at $t=0$~Myr, all of the close secondaries with $a \leq 50$~AU have experienced inward scattering. In set~1 (for which we have required at least one LMS) the region close to the host star ($\leq 50$~AU) is dominated by LMSs; they are approximately six times more populous than BDs, despite the BD-heavy initial secondary mass function (see Figure~\ref{figure:m_initial}). This distribution resembles the brown dwarf desert, i.e. the lack of BD close companions to Solar-type stars \citep{marcy2000, grether2006, kraus2008, kraus2011, sahlmann2011}. On the other hand, no such behaviour is seen in set~2, for which the initial secondary mass function is more BD-heavy. For set~2, BDs dominate the region close to the host star. A comparison between the results of set~1 and set~2 suggests that the presence of at least one LMS is therefore required for reproducing the BD desert. LMSs are routinely produced in simulation of disc fragmentation \citep[e.g.][]{stamatellos2009a}. Even if objects that form close to the host star initially have a lower mass (e.g. they start off as proto-PMOs or proto-BDs) they end up as LMSs as they accrete material form the gas-rich inner disc region.

Gravitational scattering between secondaries is accompanied by exchange of energy and angular momentum. It is therefore expected that the secondaries that experience substantial changes in their semi-major axis also obtain high eccentricities and inclinations. This is indeed the case, as shown in Figure~\ref{figure:ae}, and is qualitatively similar to the $a-e$ distributions found by \cite{forgan2015}.

The eccentricity distributions for the three classes of secondaries is shown in the top panels of Figure~\ref{figure:ae}. Although all companions were initialised on circular orbits, most of the companions remaining at $t=10$~Myr have highly eccentric ($e>0.5$) orbits, in particular the PMOs. The LMSs in set~2 have on average the smallest eccentricities, as these systems contain few secondaries massive enough to perturb these LMSs significantly. A comparison between the middle and bottom panels of Figure~\ref{figure:ae} demonstrates that it is easier for secondaries to achieve high eccentricities than it is to achieve high inclinations. The eccentricity distribution for migrated secondaries is more or less uniform, with a tendency to higher values of the eccentricity with higher degrees of migration. Interestingly almost all bound secondaries in both sets of simulations have obtained high eccentricities, even when they retain their initial orbital energy (i.e., their initial semi-major axis). Although secondaries that have migrated inwards or outwards are mostly scattered into highly inclined orbits, those remaining close to their original orbit attain much smaller eccentricities ($i\la 30^\circ$). The number of secondaries in retrograde orbits ($i>90^\circ$) is substantial, but retrograde orbits are rare in the separation range in which the secondaries were initially formed.

The cumulative inclination distribution of single and binary secondaries orbiting the host star is shown in Figure~\ref{figure:i_out}. Due to the smaller initial number of secondaries in set~2, fewer secondaries are scattered into highly-inclined orbits. Nevertheless, few bodies remain in orbits near the plane of the disc out of which they have formed. Approximately 17\% (set~1) and 5\% (set~2) of the single secondaries obtain retrograde orbits. Figures~\ref{figure:mi} and \ref{figure:ei} show a strong correlation between the orbital inclination and secondary mass, with massive secondaries (including binary secondaries and merger products) have orbits that are only mildly inclined with respect to the plane in which they originated. Binary systems orbit their host star closer to the primordial circumstellar disc than single stars. This is partially due to their larger masses, and partially due to the reduced effectivity of scattering when two secondaries pair up into one binary. Binarity among secondaries is further discussed in Section~\ref{section:binarity}.

%%%%%%%%%%%%%%%%%%%%%%%%%%%%%%%%%%%%%%%%%%%%%%%%%%%%%%%%%%%%%%%%%%%%%%%%%
%%%%%%%%%%%%%%%%%%%%%%%%%%%%%%%%%%%%%%%%%%%%%%%%%%%%%%%%%%%%%%%%%%%%%%%%%
%%%%%%%%%%%%%%%%%%%%%%%%%%%%%%%%%%%%%%%%%%%%%%%%%%%%%%%%%%%%%%%%%%%%%%%%%

\subsection{Properties of escaped secondaries} \label{section:escapers}

\begin{figure*}
  \centering\begin{tabular}{cc}
  \includegraphics[width=0.45\textwidth,height=!]{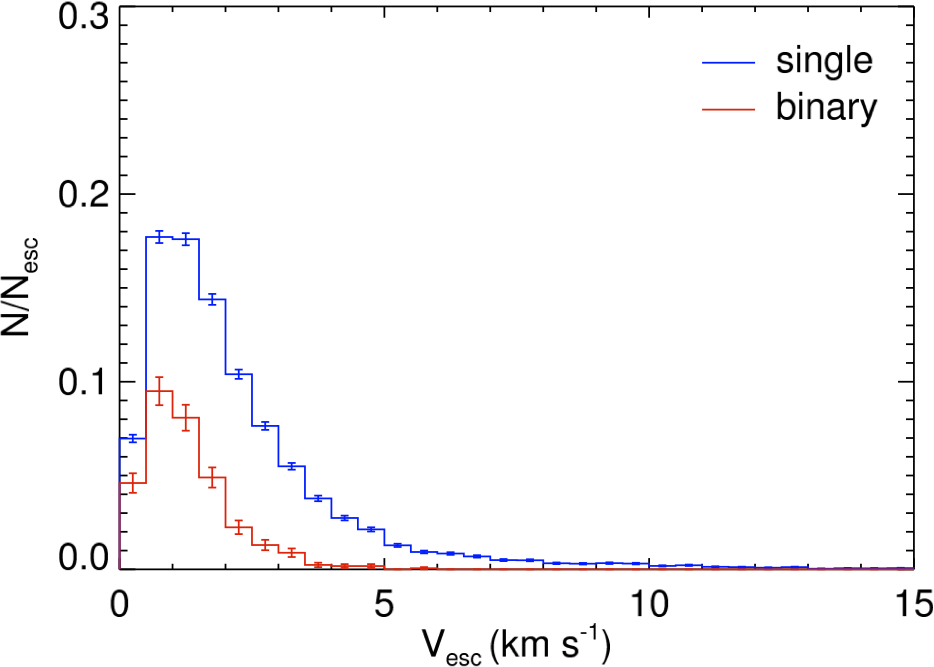} &
  \includegraphics[width=0.45\textwidth,height=!]{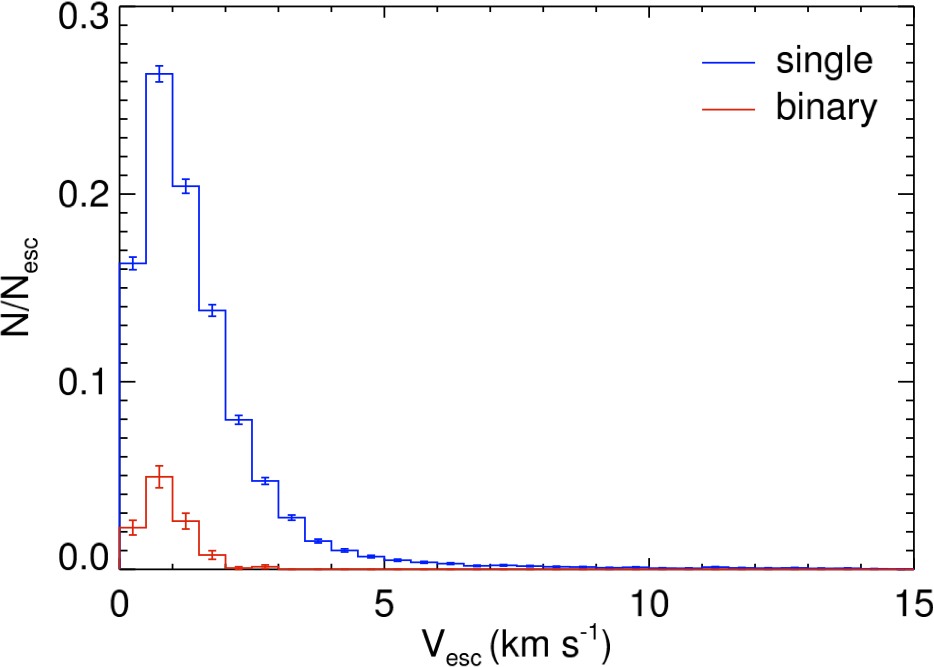} \\
  \includegraphics[width=0.45\textwidth,height=!]{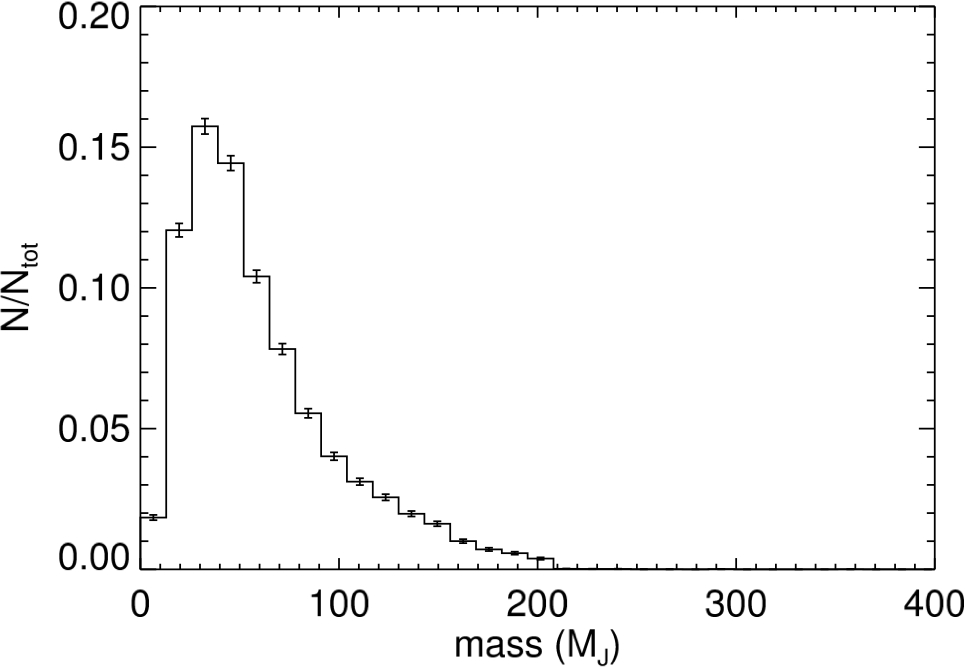} &
  \includegraphics[width=0.45\textwidth,height=!]{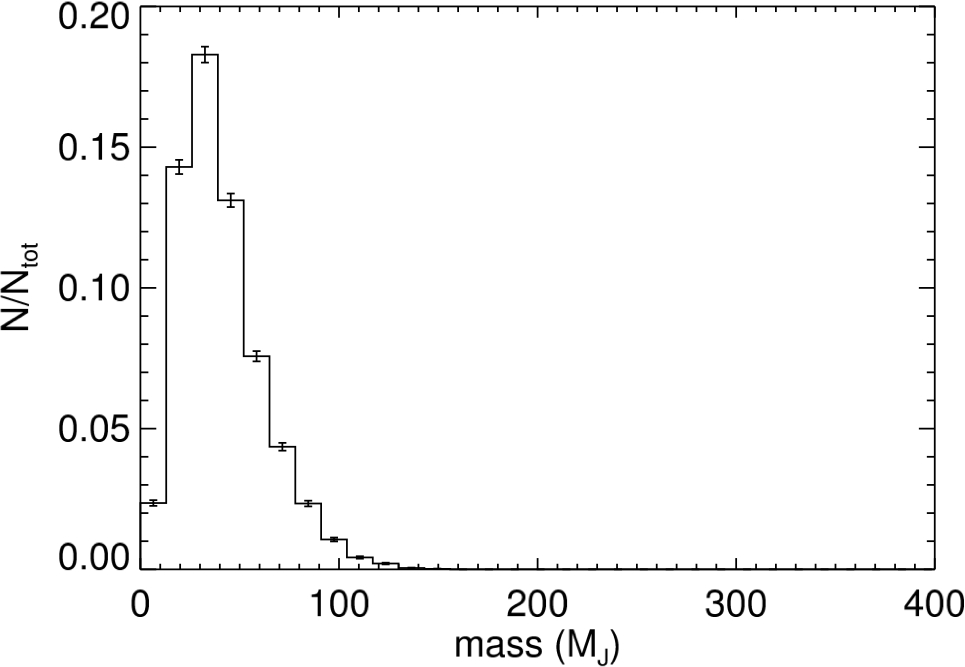} \\
  \includegraphics[width=0.45\textwidth,height=!]{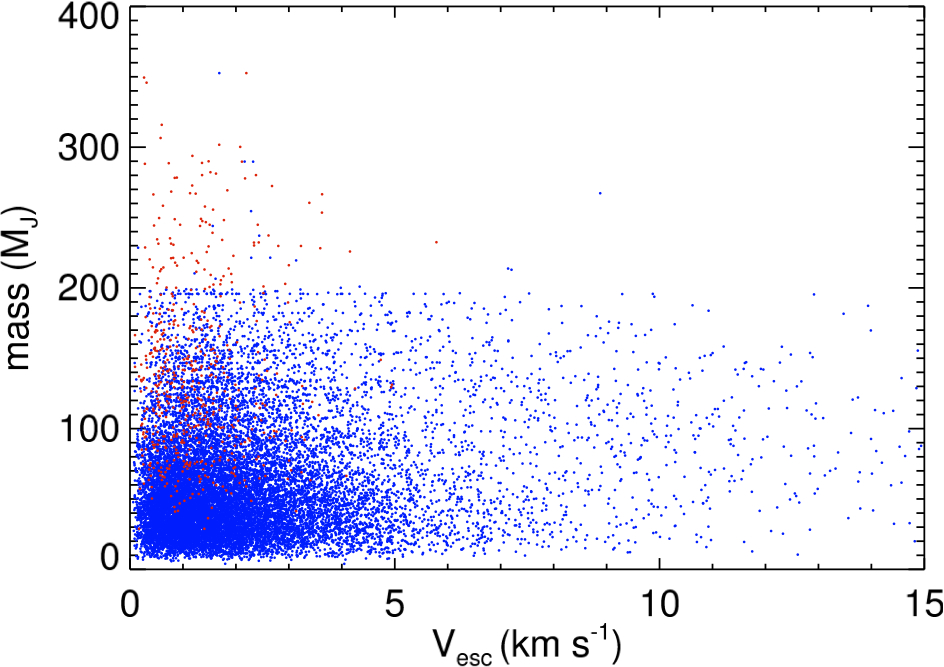} &
  \includegraphics[width=0.45\textwidth,height=!]{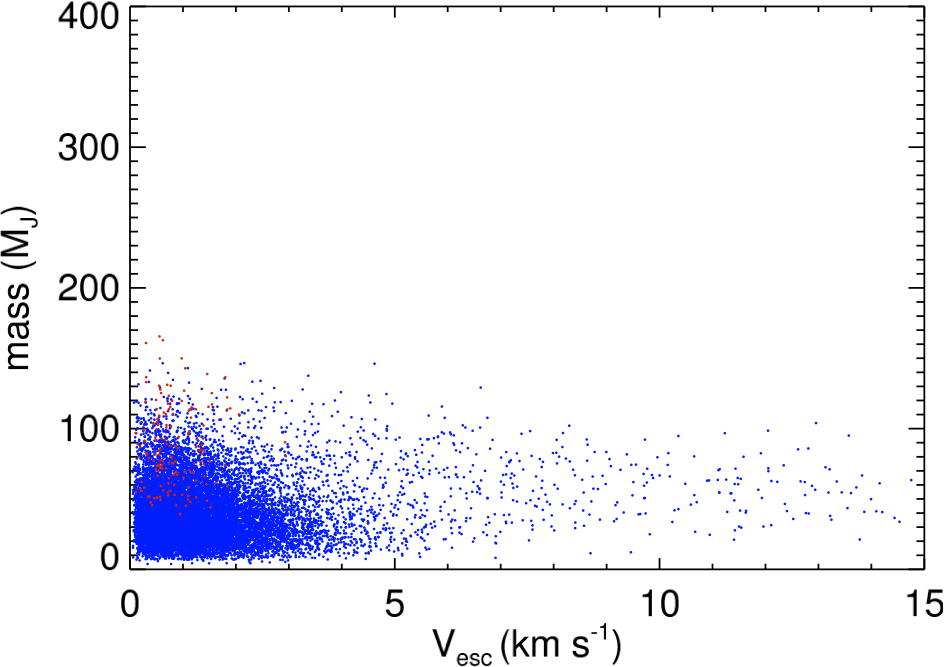} \\
  \end{tabular}
  \caption{The escape velocity distributions for single escapers (blue) and binary escapers (red) at $t=10$~Myr for set~1 ({\em top-left}) and set~2 ({\em top-right}). The red histograms are scaled up by a factor ten to allow a better comparison. The middle panels show the mass distributions of the escapers for set~1 ({\em left)} and set~2 ({\em right)}, respectively. Escape velocity versus mass at $t=10$~Myr is shown in the bottom panels for set~1 ({\em left)} and set~2 ({\em right)}. Single escapers and binary escapers are indicated with blue and red dots, respectively.
  \label{figure:v} }
\end{figure*}

Most of the secondary objects escape after gravitational scattering events. The distributions of the velocities of all escaped secondaries are plotted in the top panels of Figure~\ref{figure:v}. These velocities are calculated relative to that of the host star of the system they have escaped from. If other companions are in orbit around the host star, the host star exhibits a periodic wobble (with a time-average of zero), which in the worst case scenario (a companion of $200~\mjup$ at $a\approx 10$~AU; see Figure~\ref{figure:a_single}) results in velocity variations of 2~\kms. In the vast majority of the cases, however, these velocity variations are small. In both set~1 and set~2, most of the secondary objects attain a velocity-at-infinity smaller than 5~\kms, with a peak value at $\sim 1$~\kms. Their initial orbital velocities range between roughly 1.3~\kms{} (at 350~AU) and 3.5~\kms{} (at 50~AU). This indicates that these escapers left their initial orbit with velocity of typically less than 7~\kms{} after a scattering event.

Ejection velocities tend to be higher for set~1, which can be attributed to the larger initial number of secondary objects per system, and to a wider secondary mass spectrum, allowing the lower-mass secondary objects to be ejected with higher velocities. A tail of high-velocity escapers beyond 5~\kms{} is observed for both datasets. In total we find 27 runaway secondaries with velocities larger than 30~\kms. These include 14 BDs in set~1, one LMS in set~1, and 12 BDs in set~2. For both datasets, the fastest escapers have a velocity of almost 50~\kms. The average number of runaway ($>30$~\kms) BDs produced per system is therefore $5\times 10^{-3}$ for set~1 and $2\times 10^{-3}$ for set~2.

The middle panels of Figure~\ref{figure:v} show the mass distribution of all escaped secondaries. Lower-mass escapers are most abundant, which is partially a result of our choice for the initial secondary mass distribution. A comparison with Figures~\ref{figure:m_initial} and~\ref{figure:m_bound}, however, shows a strong preference for the ejection of low-mass secondaries. Almost all PMOs are ejected from the systems, indicating that the possibility of forming wide-orbit planetary-mass companions orbiting stars through this mechanism is small. On the other hand, the process does predict that the vast majority of these free-floating PMOs become part the Galactic field.

The correlation between the escape velocity and secondary mass is shown in the bottom panels of Figure~\ref{figure:v}. There is a preference for low-mass secondaries to be ejected with high velocity with respect to higher-mass secondaries. The objects with masses larger than $200~\mjup$ are formed through mergers at earlier times. Escaping binary systems, indicated with the red dots, tend to escape at relatively low velocities. These merger products and binary systems typically have higher masses and therefore reach relatively low velocities after a scattering event. Moreover, the fact that objects in these two categories are the "damped" combination of two primordial secondaries, their orbits are expected to be relatively stable, making it more difficult for these to be ejected.
If the systems under consideration are part of a stellar grouping (such as an open cluster), the escaped BDs and PMOs that are not immediately ejected from the cluster, gradually move to the cluster outskirts on these timescales, after which they are stripped off by the Galactic tidal field \citep[e.g.,][]{goodwin2005, wang2015, zheng2015}. 

%%%%%%%%%%%%%%%%%%%%%%%%%%%%%%%%%%%%%%%%%%%%%%%%%%%%%%%%%%%%%%%%%%%%%%%%%
%%%%%%%%%%%%%%%%%%%%%%%%%%%%%%%%%%%%%%%%%%%%%%%%%%%%%%%%%%%%%%%%%%%%%%%%%
%%%%%%%%%%%%%%%%%%%%%%%%%%%%%%%%%%%%%%%%%%%%%%%%%%%%%%%%%%%%%%%%%%%%%%%%%

\subsection{Binarity among bound and unbound secondaries} \label{section:binarity}

\begin{figure*}
  \centering
  \begin{tabular}{ccc}
  \includegraphics[width=0.315\textwidth,height=!]{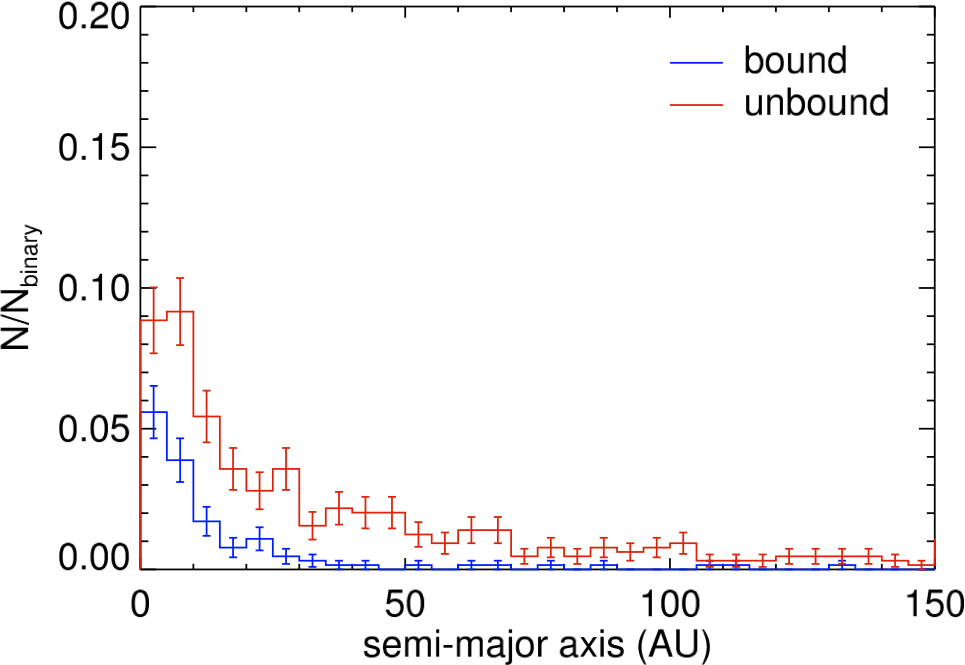} &
  \includegraphics[width=0.315\textwidth,height=!]{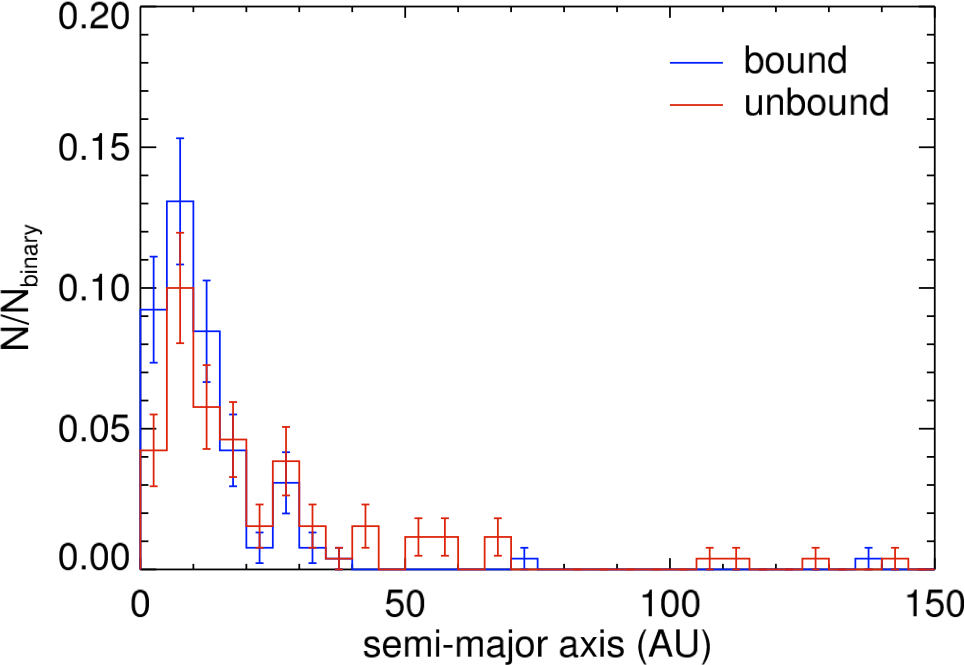} &
  \includegraphics[width=0.315\textwidth,height=!]{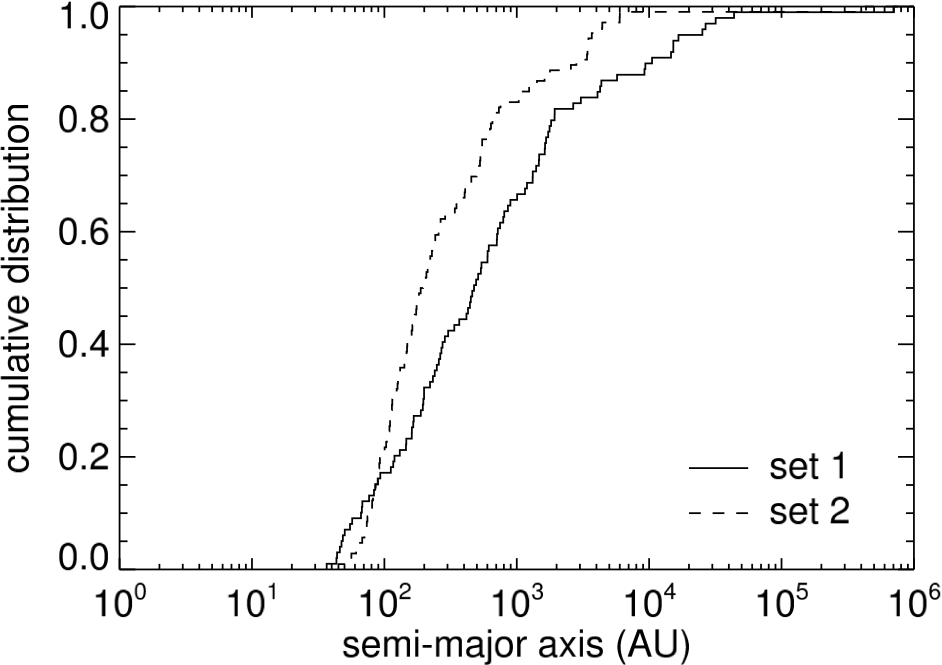} \\
  \end{tabular}
  \caption{The internal semi-major axis distributions for bound binary secondaries (blue) and unbound binary secondaries (red), for set~1 ({\em left}) and set~2 ({\em middle}), at $t=10$~Myr. The right-hand panel shows the cumulative external semi-major axis distributions of the bound binary pairs for set~1 (solid curve) and set~2 (dashed curve) at $t=10$~Myr.
  \label{figure:a_binary} }
\end{figure*}

\begin{figure}
  \centering
  \includegraphics[width=0.45\textwidth,height=!]{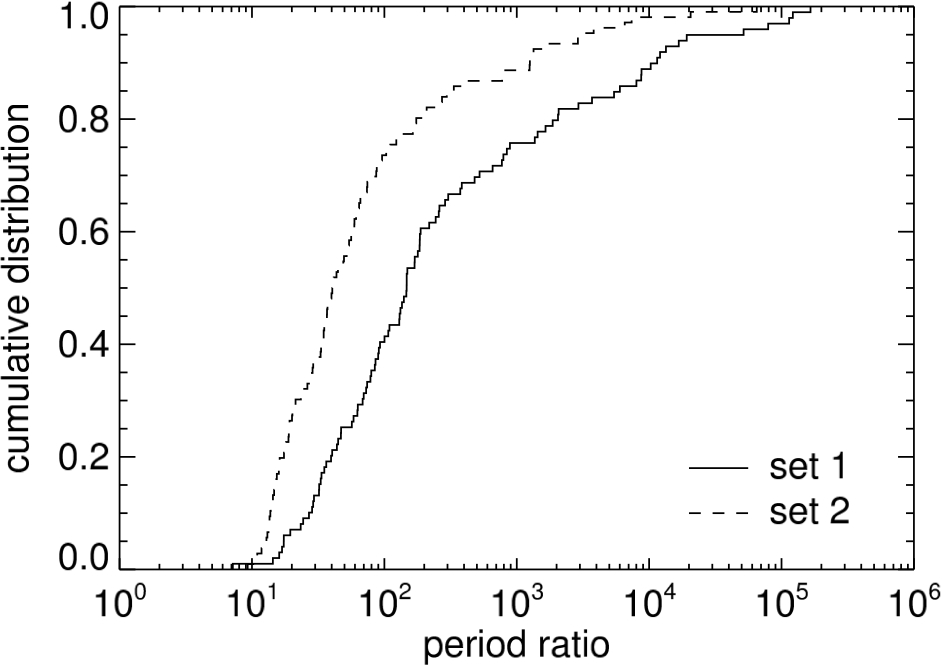} \\
  \caption{Cumulative distributions of the period ratios for the bound binary secondaries at $t=10$~Myr for set~1 (solid curve) and set~2 (dashed curve).
  \label{figure:p_binary_out_in} }
\end{figure}

\begin{figure*}
  \centering
  \begin{tabular}{cc}
  \includegraphics[width=0.45\textwidth,height=!]{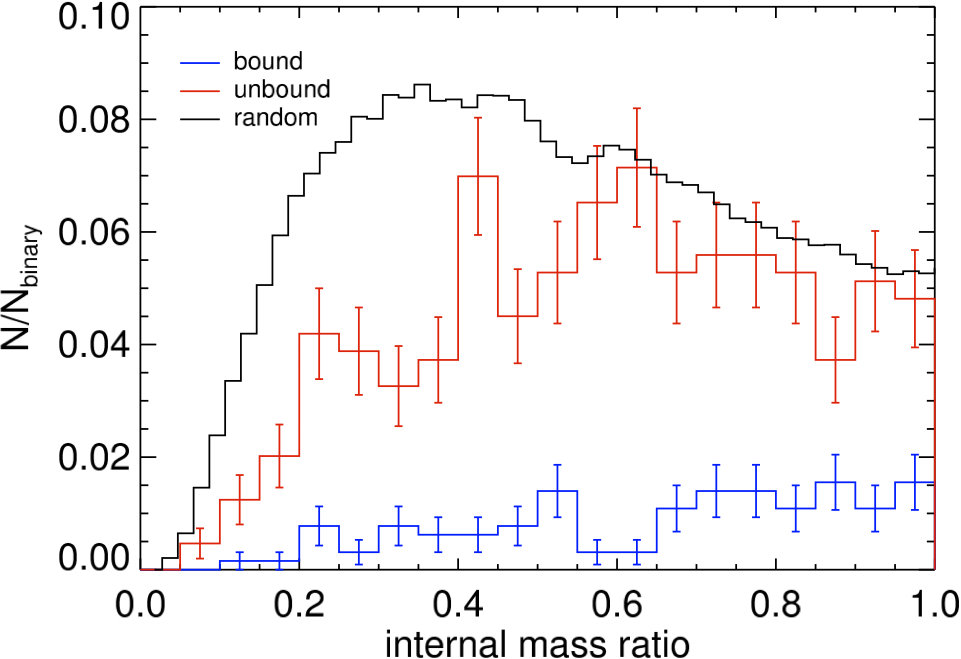} &
  \includegraphics[width=0.45\textwidth,height=!]{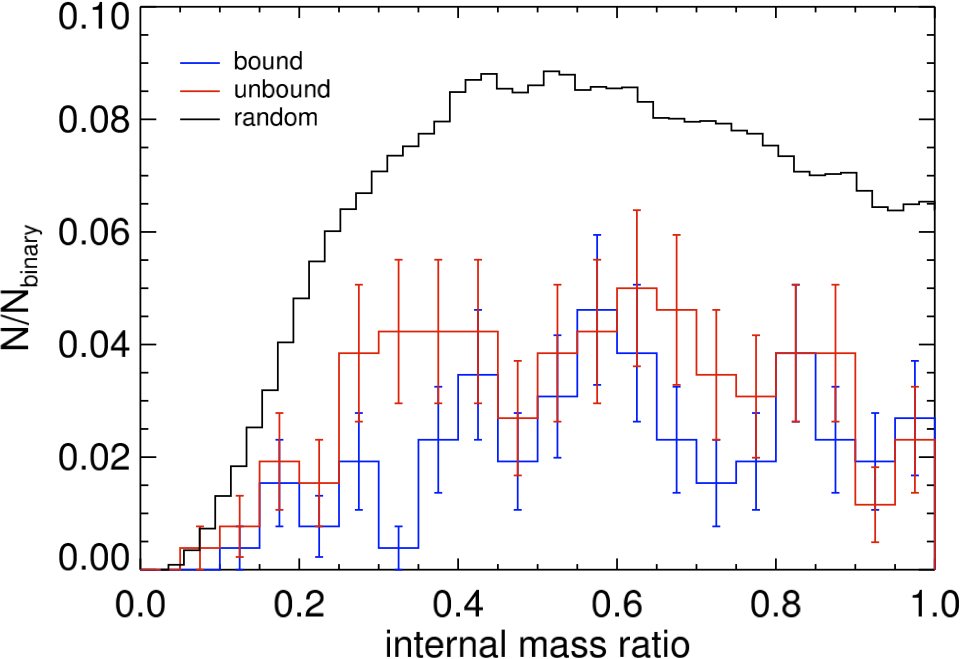} \\
  \includegraphics[width=0.45\textwidth,height=!]{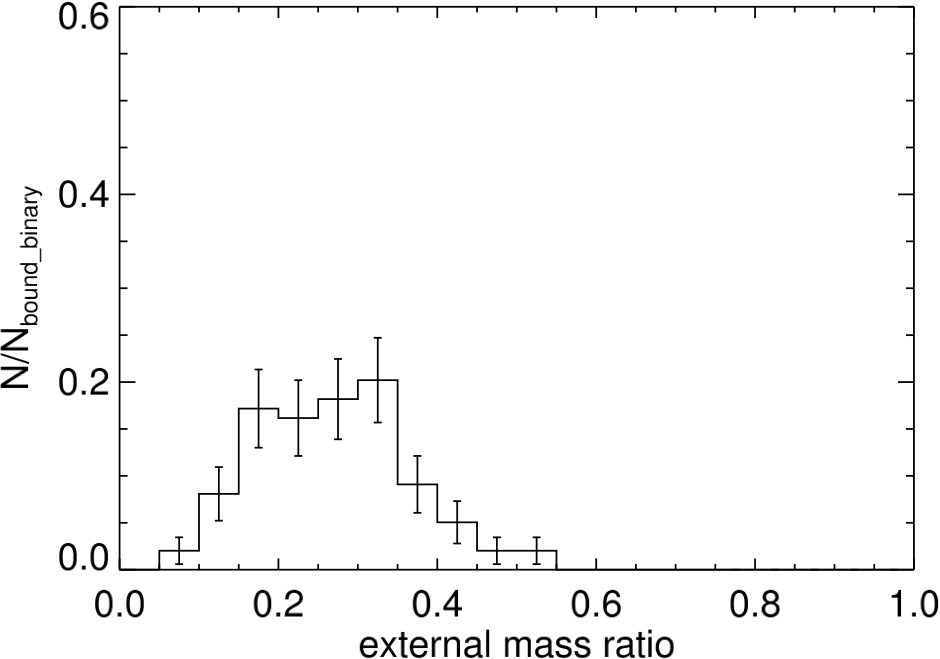} &
  \includegraphics[width=0.45\textwidth,height=!]{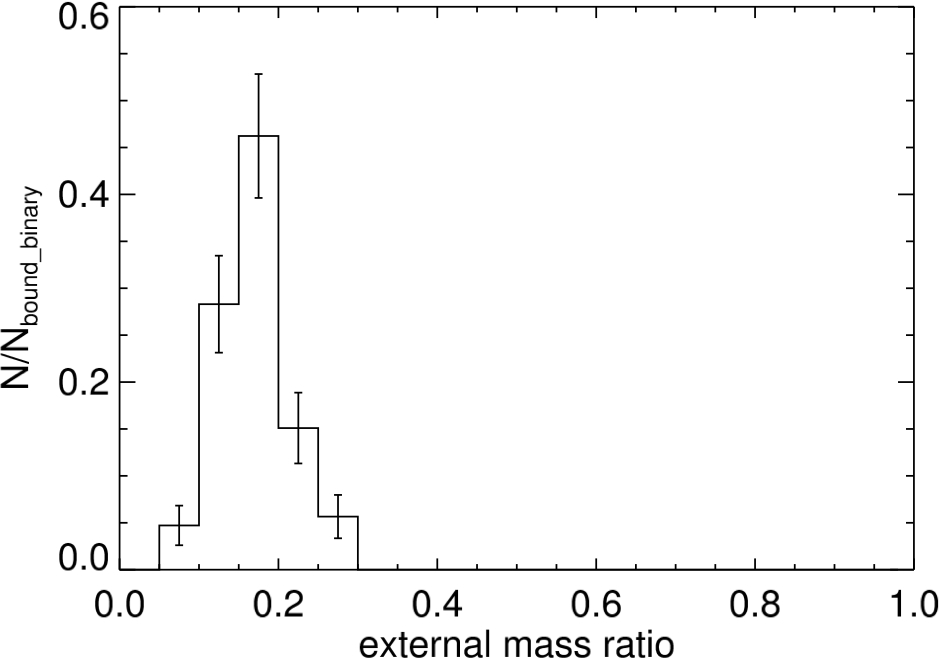} \\
  \end{tabular}
  \caption{Mass ratio distributions of the secondary binary systems at $t=10$~Myr. {\em Top:} the internal mass ratio distributions of the binary secondaries. The blue and red histograms represent bound and unbound binaries, respectively, for set~1 ({\em left}) and set~2 ({\em right}). The black histogram represent mass ratio distribution resulting from random pairing of all secondaries. {\em Bottom:} the external mass ratio distributions for bound binary secondaries, for set~1 ({\em left}) and set~2 ({\em right}). 
  \label{figure:m_ratio_in} }
\end{figure*}

\begin{figure*}
  \centering
  \begin{tabular}{cc}
  \includegraphics[width=0.45\textwidth,height=!]{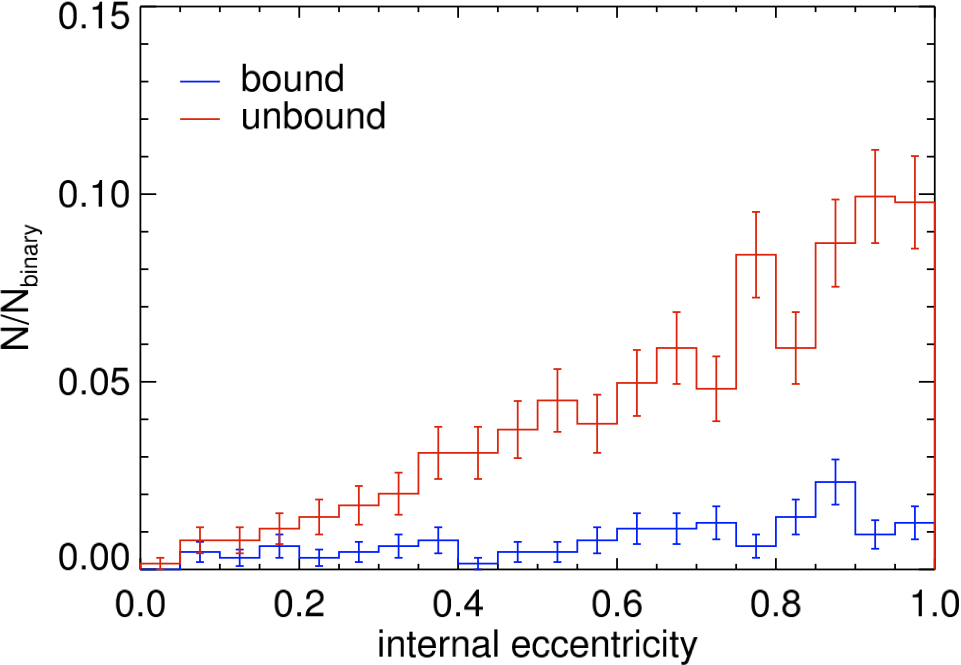} &
  \includegraphics[width=0.45\textwidth,height=!]{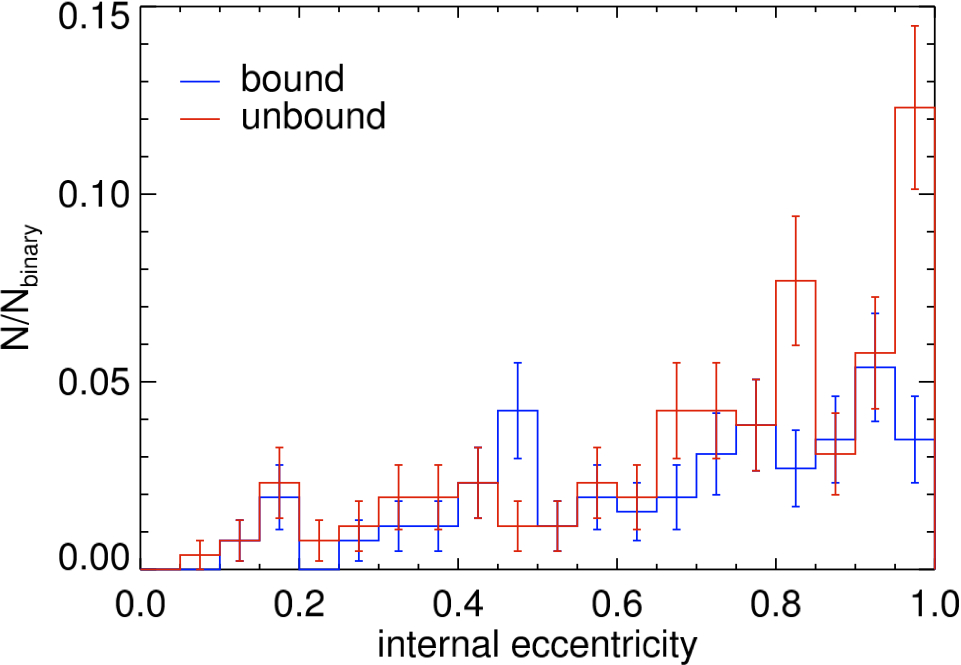} \\
  \includegraphics[width=0.45\textwidth,height=!]{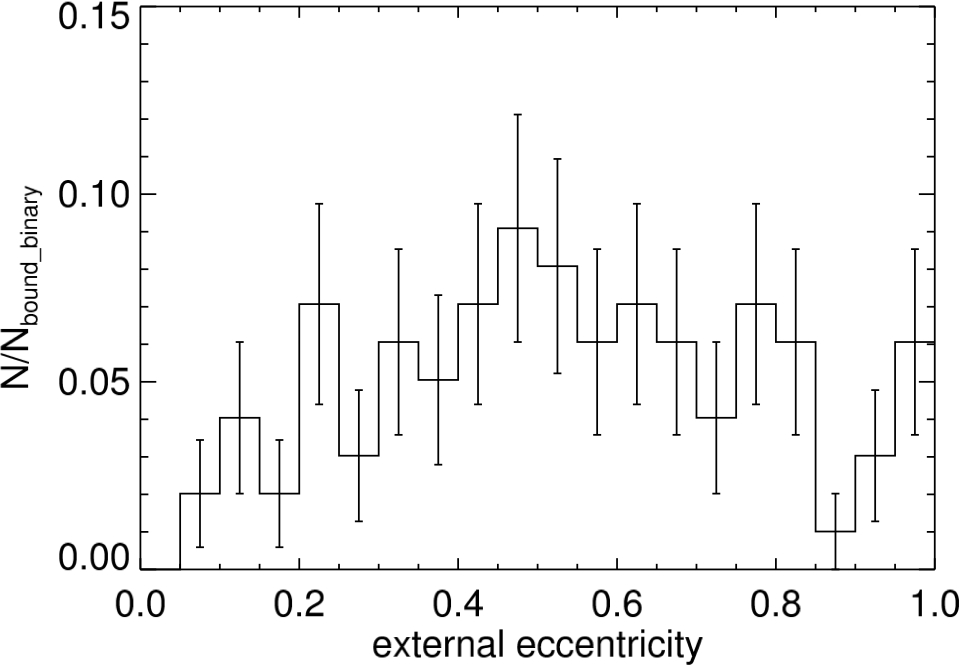} &
  \includegraphics[width=0.45\textwidth,height=!]{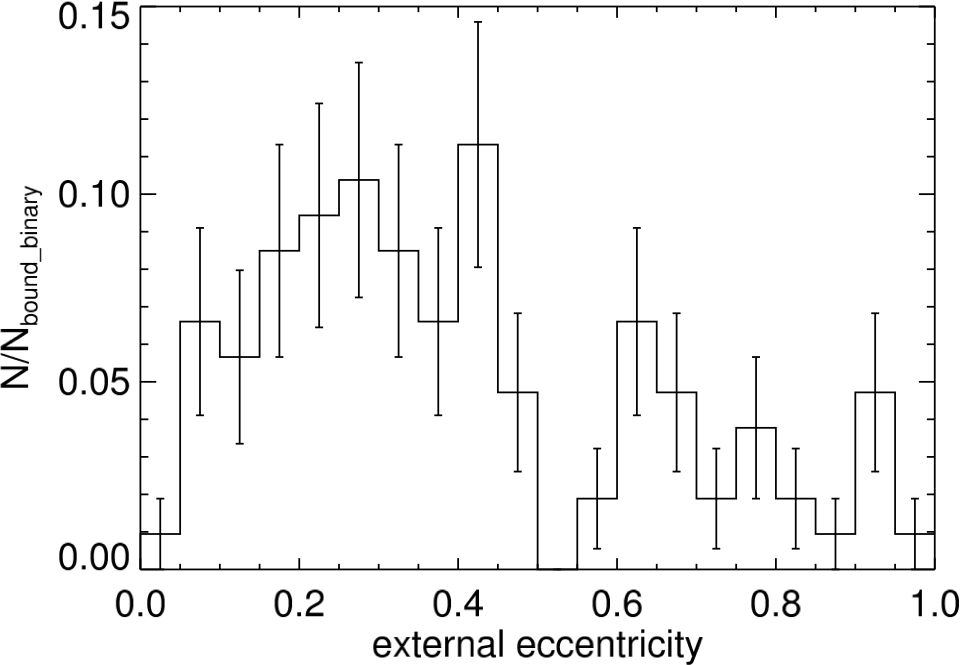} \\
  \end{tabular}
  \caption{Eccentricity distributions of the binary secondaries at $t=10$~Myr, for set~1 ({\em left}) and set~2 ({\em right}). {\em Top:} the internal eccentricity distributions for the bound (blue) and unbound (red) binaries.  {\em Bottom:} the external eccentricity distributions. 
  \label{figure:e_binary_in} }
\end{figure*}

\begin{figure}
  \centering
  \begin{tabular}{ccc}
  \includegraphics[width=0.315\textwidth,height=!]{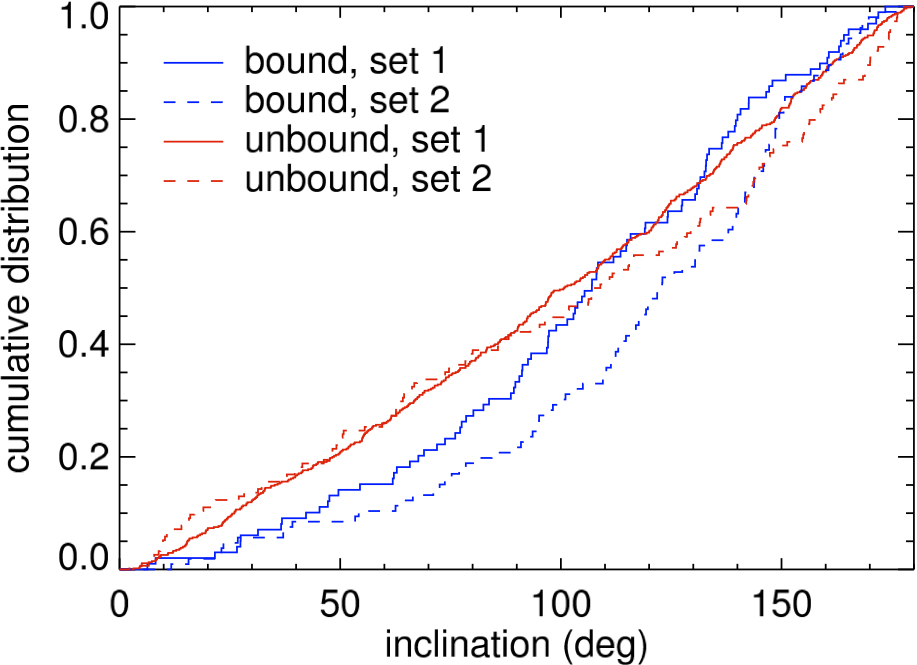} &
  \includegraphics[width=0.315\textwidth,height=!]{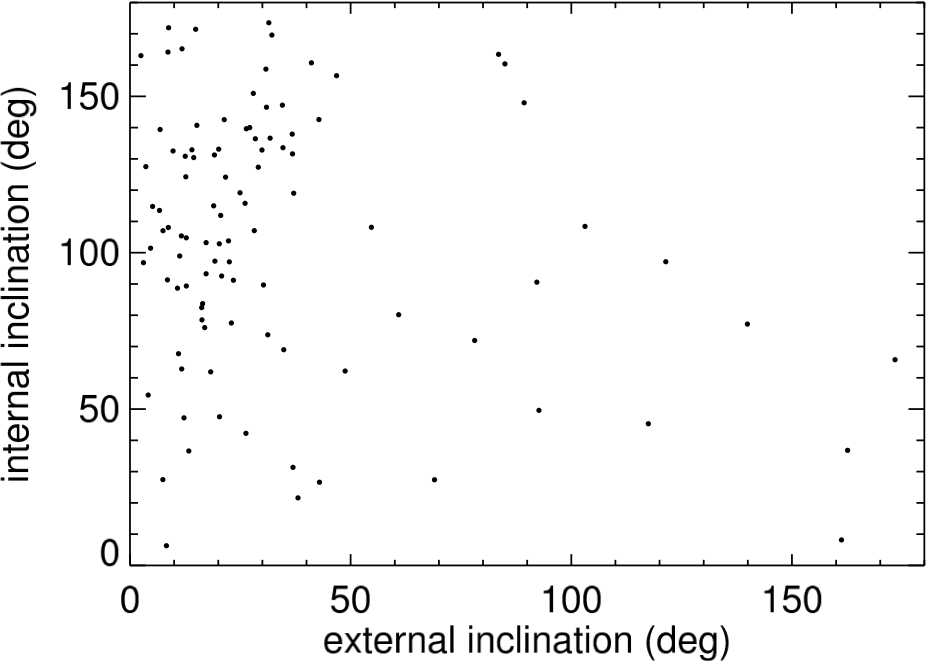} &
  \includegraphics[width=0.315\textwidth,height=!]{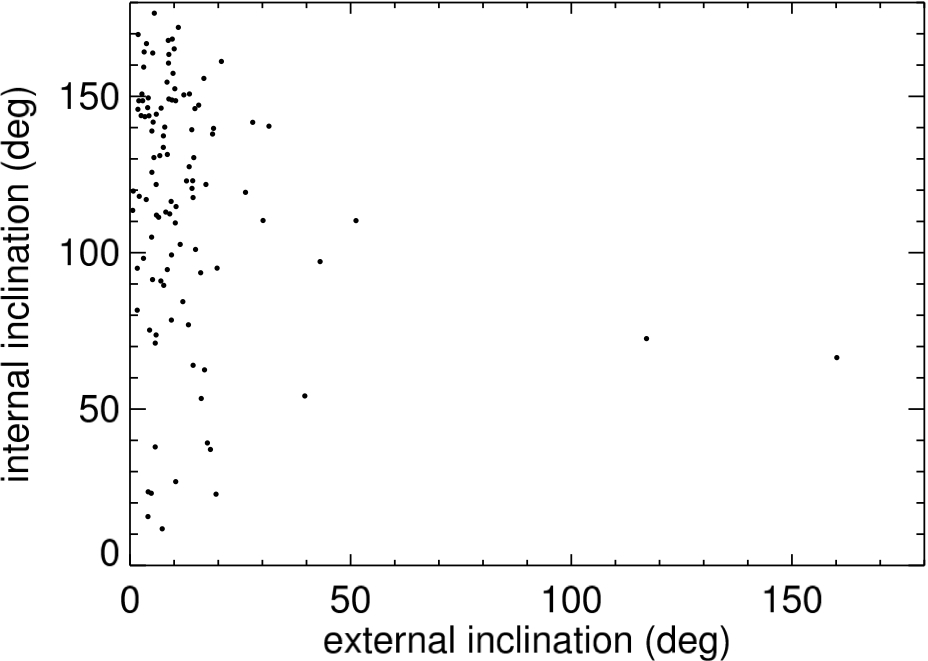} \\
  \end{tabular}
  \caption{Inclination distributions of the binary secondaries at $t=10$~Myr. {\em Left:} cumulative internal inclination distributions for bound (blue) and unbound (red) binaries in set~1 (solid curves) and set~2 (dashed curves). {\em Middle:} internal versus external inclination for the bound binary secondaries in set~1. {\em Right:} internal versus external inclination for the bound binary secondaries in set~2.
  \label{figure:i_binary_in} }
\end{figure}

For very low-mass stars and brown dwarfs, the observed companion fraction ranges between 10\% and 30\% \citep[e.g.,][]{bouy2003, reid2008, joergens2008, goldman2008}. As observational techniques become more accurate, more and more binary systems are found to be of higher multiplicity, both in embedded and young star forming regions \citep[e.g.,][]{kraus2007, connelley2008a, connelley2008b, kraus2011} and in the older stellar populations \citep[e.g.,][]{tokovinin2006, correia2006, eggleton2008, law2010, allen2012, burgasser2012, tokovinin2014a, tokovinin2014b}. These systems may have originated from different mechanisms, such as decay of disc-fragmented systems \citep[this paper; see also][]{podsiadlowski2010}, or from the decay of non-hierarchical triple systems \citep[e.g.,][]{umbreit2005, reipurth2015}.
In our models, a fraction of the secondary objects pair into binaries. These binary secondaries either orbit the host star or escape from the system and become very low-mass binaries in the Galactic field. We define the binary fraction $F=B/(S+B)$, where $S$ is the number of single secondaries, and $B$ the number of binary secondaries. The binary fraction for several sub-populations of secondaries are listed in Table~\ref{table:leftover}.

The external semi-major axes of the binary secondaries orbiting the host star are shown as red dots in Figures~\ref{figure:a_single} and~\ref{figure:ae}, and the internal semi-major axis distributions for both the bound and escaped binary secondaries in Figure~\ref{figure:a_binary}. The internal semi-major axis distributions of the bound and escaping binary secondaries are very similar, although there are fewer bound binaries with wide ($\ga 30$~AU) orbits, due to gravitational perturbations of the host star and other companions. The internal semi-major axes are well within the radius of the Hill sphere of the two components and therefore guarantee longevity when no other secondary objects approach the binary. 
Bound and escaping binaries are wider for set~2 (peak at $\sim 10$~AU) than for set~1 (peak at $\sim 5$~AU). This difference is a result of the initially larger number of companions in set~1 and the subsequent destruction of wide binary pairs by other secondaries.
For set~1, there are more escaped binaries than bound binaries at any semi-major axis, hinting at disruption of bound binary pairs by other remaining companions. For set~2, on the other hand, the number of bound and escaped binaries is roughly equal, although the majority of the tight ($<5$~AU) binaries is bound. 
Observationally, the separation distribution of very low mass binary systems is still poorly constrained due to the resolution limit of imaging surveys, through which most companions are detected \citep[e.g.,][]{burgasser2007b}. The semi-major axes distributions for the very low mass binaries in our model are in reasonable agreement with the observations of \cite{maxted2005}, \cite{burgasser2005}, \cite{burgasser2007a}, \cite{bergfors2010} and \cite{bardalez2014a, bardalez2014b}.

The external semi-major axes of the bound binary secondaries are substantially larger than the internal semi-major axes, as can be seen in Figure~\ref{figure:a_binary}. This large ratio guarantees long-term stability. Figure~\ref{figure:p_binary_out_in} shows the distributions of $P_{\rm ext}/P_{\rm int}$ for the bound binary systems, where $P_{\rm int}$ is the orbital period of the two secondaries around their mutual centre of mass, and $P_{\rm out}$ is the orbital period of the binary secondary around the host star. Although systems with period ratios larger than a factor ten can be stable, most multiple systems have period ratios of tens to thousands. These large ratios guarantee stability for long periods of time. The hierarchical triple systems in set~2 are substantially more compact than those in set~1, due to the lower frequency of perturbing secondaries and the smaller magnitude of these perturbations (as the typical companion masses are lower in set~2).

As mentioned above, in several systems two secondaries form a binary system that is either bound to the host star or escapes. In these cases, we define the internal mass ratio $q_{\rm int}$ and external mass ratio $q_{\rm ext}$ as
\begin{equation}
q_{\rm int} = \frac{m_2}{m_1} \quad\quad{\rm and}\quad\quad q_{\rm ext}=\frac{m_1+m_2}{M} \quad\quad ,
\end{equation}
where $M=0.7~\msun$ is the mass of the host star, and $m_1$ and $m_2$ the masses of the two companions, where $m_1 \ge m_2$. The extremes that these quantities can possibly obtain are set by our initial conditions in Table~\ref{table:initialtable}, $0.005\leq q_{\rm int} \leq 1$ and $0.003\leq q_{\rm ext} \leq 0.55$. It should be noted that these limits are rarely reached as they involve combining either the lowest-mass secondaries or the highest-mass secondaries into binaries. Moreover, the limits may change slightly when physical collisions are taken into consideration.
The internal and external mass ratio distributions for all bound and escaping binary secondaries are shown in Figure~\ref{figure:m_ratio_in}. All internal mass ratio distributions are biased towards larger $q_{\rm int}$ than what is expected for gravitational random pairing of secondaries \citep[see][for an extensive discussion on the different pairing functions of binary components]{kouwenhoven2009}. The frequency of low ($q_{\rm int}<0.1$) mass ratio binaries is small as it is unlikely to have a chance interaction between two bodies of very different mass, as a large majority of the secondaries in our initial models have masses in the range $20-100~\mjup$. Binaries $q_{\rm int}>0.8$ are somewhat more common than what is expected from random pairing, which is mainly a result of the fact that low-mass companions are more easily ejected than captured during scattering events. It should be noted that the internal mass ratio distributions for bound binaries and escaping binaries are very similar. The external mass ratio distribution is limited to $q_{\rm ext}\la 0.55$ for set~1 and $q_{\rm ext}\la 0.3$ for set~2. These upper limits are a result of our initial conditions, including the initial secondary mass distribution.
Despite the large number of surveys, the mass ratio distribution among very low-mass stars and brown dwarfs is still poorly understood. Only a few studies involve direct dynamical mass measurements \citep[e.g.,][]{konopacky2010}. Adaptive-optics imaging surveys are more common and allow measurements of photometric masses. The imaging survey of 16 objects in the Hyades cluster by \cite{duchene2013b} and of 124 field M-type dwarfs by \cite{bergfors2010} suggests that the BD mass ratio distribution may not be as peaked to unity as previously thought, and are therefore in reasonable agreement with the results of our simulations.

The internal eccentricity distributions and external eccentricity distributions for our models are shown in Figure~\ref{figure:e_binary_in}. The internal eccentricity distributions $f(e_{\rm int})$ of the bound and escaped binary secondaries in both sets are roughly thermal \citep[][]{heggie1975}, following the proportionality $f(e_{\rm int})\propto e_{\rm int}$, although the escaped binaries have higher internal eccentricities as they do not experience any additional evolution after formation in the absence of gravitational perturbations by the host star and other remaining secondaries. The external eccentricity distributions $f(e_{\rm ext})$ of the bound binary secondaries are broad, with few nearly circular orbits (related to their chaotic formation process) and few high-eccentricity orbits (due to dynamical evolution in the presence of the host star). The internal and external eccentricity relation for bound binaries are independent.
Observationally, little is known about the eccentricity distribution for very low-mass binaries, and have mostly been limited to time-consuming surveys among relatively tight binaries \citep[see, e.g.,][]{dupuy2011, konopacky2010}, and should therefore not be directly compared to our results.

In the case of a bound or escaping pair of secondaries orbiting a common centre-of-mass, we define the {\em internal} inclination as the angle between their orbital plane and the plane of the primordial disc. The {\em external} inclination is the angle between the orbit of the centre of mass around the host star and the plane of the primordial disc.
The cumulative external inclination distribution of the secondary objects at $t=10$~Myr and the correlation between external inclination and mass were discussed in Section~\ref{section:elements} (see Figures~\ref{figure:i_out} and~\ref{figure:mi}). The correlation between binary mass and external inclination follows the same trend as that of the single companions, although it should be noted that binaries tend to be more massive, and as a result, tend to have smaller inclination.
As set~2 has fewer and lower-mass secondary objects, scattering events are rarer, and there are fewer secondaries with high inclinations than in set~1. Although the initial inclination distribution is uniform between $0^\circ$ and $5^\circ$, the final distribution covers all values. The cumulative internal inclination distributions $f(i_{\rm int})$ of the bound and escaped binaries, as well as the correlations between the internal and external inclinations of the bound binary secondaries are plotted in Figure~\ref{figure:i_binary_in}. For a set of randomly oriented orbits the inclinations distribution is $\frac{1}{2}(1-\cos i_{\rm int})$, and 50\% of the orbits have inclinations less than $90^\circ$. In all distributions in the left-hand panel of Figure~\ref{figure:i_binary_in}, however, less than half of the orbits have inclinations less than $90^\circ$. The internal inclination distributions of the escaping binary systems are more or less flat, while bound binaries have a distribution that is strongly biased toward retrograde orbits: roughly 65\% and 80\% of the bound binary secondaries have a retrograde orbit with respect to the circumstellar disc out of which they have formed, although most of these retrograde binaries have a prograde orbit around the host star. Binary systems with a lower external inclination tend to have a higher internal inclination, although this correlation is not strong. Binaries with higher external inclinations tend to have higher external eccentricities, as for the single bound secondaries that were previously discussed in Figure~\ref{figure:ei}.

%%%%%%%%%%%%%%%%%%%%%%%%%%%%%%%%%%%%%%%%%%%%%%%%%%%%%%%%%%%%%%%%%%%%%%%%%
%%%%%%%%%%%%%%%%%%%%%%%%%%%%%%%%%%%%%%%%%%%%%%%%%%%%%%%%%%%%%%%%%%%%%%%%%
%%%%%%%%%%%%%%%%%%%%%%%%%%%%%%%%%%%%%%%%%%%%%%%%%%%%%%%%%%%%%%%%%%%%%%%%%

\section{Discussion} \label{section:discussion}

The most obvious question to ask is whether there is any observational evidence of the sort of (massive) disc fragmentation we simulate in this paper having actually
occurred.  This is actually a very difficult question to answer. Ideally, one would like to observe a massive gas disc, however they are extremely short-lived, and so the chance of observing a massive disc is small.  Therefore, in this paper we simulate the outcome of evolution of the fragments in order to examine possible observable signatures. The disc fragmentation mechanism simulated here, based on the SPH simulations of \cite{stamatellos2009a} is not universal. That is, they are not expected to occur all of the time in all environments. Therefore, any comparison with observations will include, at best, a mixture of the products of this mechanism and possibly many others.

It is worth noting that this disc fragmentation mechanism requires intermediate to low star formation densities to operate.  Although massive discs only live a short time before fragmenting, they must be able to form in the first place, which seems unlikely at densities exceeding a thousand stellar systems per cubic parsec, similar to those simulated by, e.g., \cite{bate2009}, in which dynamical interactions constantly perturb and truncate discs. These perturbations are usually destructive \citep[e.g.,][]{olczak2006, olczak2008}, although they can under specific circumstances induce disc fragmentation \citep[e.g.,][]{thies2005, thies2010, thies2015}. If this is a typical environment for star formation \citep[as argued by][]{marks2011} then this mechanism is probably fairly rare, occurring only in rare diffuse star forming regions such as Taurus. On the other hand, if star formation often occurs at relatively low density \citep[see, e.g.,][]{allison2009, bressert2010, king2012, parker2014, wright2014} then this mechanism might be relatively common. Even if the disc fragmentation mechanism is relatively common we must ask how common. Again, this is unclear, but if this occurs in (say) 20 per cent of stars in regions of low to intermediate density, we might expect to see the outcome of the dynamical evolution in a significant number of
systems. If it occurs in only a few per cent of stars, the remnants will be rare. In addition, the disc fragmentation mechanism cannot work around all stars.  To build a large, massive disc which is able to fragment the host star must be single, or a close binary (to build a circumbinary disc), or a wide enough binary to build a massive circumprimary disc.  Many systems with relatively massive companions at $10-100$~AU will be unable to form a massive disc.

As we argued above, these massive discs are unlikely to form if the star formation process occurs at high density (i.e., stars interact and know about each other whilst accumulating mass and building discs).  At moderate to low densities massive discs might form and be able to fragment in the way seen by, e.g., \cite{stamatellos2007b} and \cite{stamatellos2009a}.  However, even in environments where disc fragmentation can produce the systems modelled in our study, subsequent star cluster dynamics may affect the systems, such that they may evolve differently from the isolated systems described in this paper. At moderate densities (tens to hundreds of stars per cubic parsec), or if stellar densities increase after formation \citep{allison2009} then dynamical interactions between systems will become important and dynamically alter the properties of multiple systems \citep[e.g.,][]{aarseth2003, spurzem2009, malmberg2011, hao2013}. Therefore, the best places to look for the types of system we simulate here are local low density star forming regions which we believe have always been at low density.  The best candidate is Taurus which is low density, it is also substructured suggesting it is dynamically young and so has always been at low density \citep[see][]{parker2014}, and is close enough to have been well studied down to brown dwarf masses.  Unfortunately, Taurus has relatively few systems ($\sim 300$) and so statistics will be poor.

The simulations in this paper suggest that massive disc fragmentation will produce a population with relatively high-order multiplicity with low-mass stars concentrated at $10-10^3$~AU, brown dwarfs widely spread between 10~AU and $10^4$~AU, and planetary-mass objects at $>500$~AU (see, e.g., Figure~\ref{figure:a_single}). In observations, these populations will be mixed with other populations which formed $10-100$~AU more massive multiples. Interestingly, Taurus does seem to have some wide planetary mass companions to stars \citep[][]{bowler2014, kraus2014}, and a large number of high-order multiples including at least two sextuples \citep{kraus2011}.  These systems might have been formed via massive disc fragmentation (certainly to produce sextuple systems a significant amount of fragmentation must have occurred and a massive disc seems a good candidate for this).

In summary, it is currently probably impossible to say how important massive disc fragmentation is in star formation.  Firstly, it is unclear what fraction of systems are able to build a massive disc to undergo disc fragmentation (and this is probably environment-dependent).  Secondly, observations tend not to be deep or sensitive enough to detect planetary mass companions and/or close low-mass star companions except in very nearby regions where the statistics is poor.  Thirdly, the outcome of massive disc fragmentation is varied (as it is a chaotic process).  Fourthly, any products of massive disc fragmentation can be further processed by dynamical interactions with other bodies in their environments.  And finally, any `intact' products of massive disc fragmentation are mixed with processed systems and systems which did not undergo massive disc fragmentation in unknown proportions.

%%%%%%%%%%%%%%%%%%%%%%%%%%%%%%%%%%%%%%%%%%%%%%%%%%%%%%%%%%%%%%%%%%%%%%%%%
%%%%%%%%%%%%%%%%%%%%%%%%%%%%%%%%%%%%%%%%%%%%%%%%%%%%%%%%%%%%%%%%%%%%%%%%%
%%%%%%%%%%%%%%%%%%%%%%%%%%%%%%%%%%%%%%%%%%%%%%%%%%%%%%%%%%%%%%%%%%%%%%%%%

\section{Conclusions} \label{section:conclusions}

We have carried out a study on the dynamical evolution of systems formed through the gravitational fragmentation of circumstellar discs, and study the dynamical properties of the, mostly brown-dwarf mass, circumstellar objects. We have carried out two ensembles of simulations of systems, which we refer to set~1 and set~2, all of which have a host star of mass $0.7~\msun$ (see Table~\ref{table:initialtable} for details). Companions have masses of $1-200~\mjup$ and are referred to as planetary-mass objects (PMOs), brown dwarfs (BDs) and low-mass hydrogen-burning stars (LMSs), depending on their mass. The initial conditions in set~1 correspond to the outcome of the SPH simulations of \cite{stamatellos2009a}. Set~2 represents systems formed from lower-mass circumstellar discs. Our main results can be summarised as follows:
\begin{enumerate}

\item At $t=10$~Myr, approximately 22\% and 9.8\% of the host stars are single stars in set~1 and set~2, respectively, while roughly 47\% of the host stars in both sets have one companion left. Systems with two remaining secondaries are relatively common (25\% and 42\%, respectively), while only 4\% of the host stars in set~1 and 2\% of the host stars in set~2 have three companions left. Systems of higher-order multiplicity still exist at $t=10$~Myr but are rare.

\item Approximately half of the secondary objects are ejected from their host star (or experience a physical collision) within 20\,000~yrs for set~1 and within 400\,000~yrs for set~2. This difference can attributed mainly to the smaller initial number of secondaries in set~2. For both models, approximately 7\% of the secondaries escape between 1~Myr and 10~Myr. At 10~Myr, most systems have relaxed, and further decay beyond this is minimal. 
 
\item The majority of the secondaries escape from their host star within 10~Myr. Being the most massive secondaries, LMSs have most probability of remaining bound to their host star, with probabilities ranging from 36\% (set~1) to 79\% (set~2). For BDs these values are reduced to 7.2\% and 28\%, respectively. PMOs are most easily ejected, with only $2-5\%$ remain part of the system in which they formed.

\item At $t=10$~Myr, the average number of very-low mass binary secondaries formed per disc-fragmented system is 0.21 for set~1 and 0.04 for set~2. Out of these, 15\% (set~1) to 41\% (set~2) remain bound to the host star, in a stable multiple configuration. At the end of the simulations, $2-3\%$ of the systems form hierarchical multiple systems, where two companions have paired up into a binary secondary that orbits the host star. The majority ($65-80\%$) of these pairs have retrograde internal orbits, and the majority of this pairs have a prograde orbit around the host star. These bound pairs most have internal semi-major axes smaller than 20~AU, a relatively flat internal mass ratio distribution, and a more or less thermal internal eccentricity distribution.

\item Secondaries are mostly ejected from their host systems with terminal velocities smaller than 5~\kms. Several high-velocity escapers $>30$~\kms{} are generated after violent encounters with other secondaries, including runaway BDs with velocities up to almost 50~\kms. 
 Among all secondaries in the model, $1.3-4.9\%$ escape from their host star as part of binary pairs, providing a possible formation mechanism for very low-mass binary systems in the Galactic field. Their semi-major axes are typically smaller than 50~AU, and are somewhat wider than those of the bound secondary pairs. They have approximately a flat mass ratio distribution,  a thermal eccentricity distribution, and high internal inclinations. 

\item Physical collisions involving the host star and/or secondaries are common. In our simulations we find (per host star) on average $0.04-0.08$ physical collisions between secondaries, the majority of which occur within several orbital periods. Collisions with the host star are more common ($0.18-0.43$ per host star), and occur on a timescale of a thousand orbital periods, i.e., typically a million year.

\item Gravitational scattering between companions results in a strong evolution of the orbital elements. At $t=10$~Myr many companions have migrated inwards or outwards, and the semi-major axis range in which they had formed is mostly evacuated. The changes in semi-major axis, eccentricity, and inclination tend to be larger for less massive companions. Within the limits of our initial conditions of set~1, our results are in good agreement with the observed brown dwarf desert.

\item Almost all planetary-mass objects escape from their host star within 10~Myr and become free-floating objects. Only $2-5\%$ of the planetary-mass objects remain bound to the host star, generally on very wide orbits ($a > 500$~AU) at the end of the simulations. Gravitational disc fragmentation is therefore not a viable way to form hot Jupiters. The final configuration often contains secondaries with (hypothetical) Jupiter-crossing orbits (and sometimes Earth-crossing orbits). This implies that it is not likely that planet formation through the core-accretion scenario will occur after dynamical relaxation of the system has completed.

\end{enumerate}

Throughout this paper we have focused on the outcome of the fragmentation of massive discs surrounding stars of mass $M = 0.7~\msun$. The properties of objects formed through disc fragmentation will be different for host stars of different masses, and for circumstellar discs of different masses. In addition, we have neglected the presence of small amounts of gas left-over in the discs around the host stars in our simulations. Future work on the long-term evolution of these systems with the inclusion of remnant gas may be necessary, for example using the SEREN code \citep[e.g.,][]{hubber2011}, or the AMUSE framework \citep{portegies2013, pelupessy2013, cai2015}.

We have limited our study to evolution during the first 10~Myr of the disc-fragmented systems. At this time, approximately 30\% of the systems in set~1 and 44\% of the systems in set~2 are multiple. Although most of these systems are stable for much longer, a fraction of these will decay over billions of years. The results presented in this paper are therefore useful for the study of young stellar populations. In order to carry out a proper comparison with the observed low-mass stellar population in the Solar neighbourhood, the systems have to be integrated over much longer timescales; this will be the topic of our future work.

%%%%%%%%%%%%%%%%%%%%%%%%%%%%%%%%%%%%%%%%%%%%%%%%%%%%%%%%%%%%%%%%%%%%%%%%%
%%%%%%%%%%%%%%%%%%%%%%%%%%%%%%%%%%%%%%%%%%%%%%%%%%%%%%%%%%%%%%%%%%%%%%%%%
%%%%%%%%%%%%%%%%%%%%%%%%%%%%%%%%%%%%%%%%%%%%%%%%%%%%%%%%%%%%%%%%%%%%%%%%%

\section*{Acknowledgments}

We thank the anonymous referee for insightful and useful comments that helped to improve the paper.
We would like to express our gratitude to Sverre Aarseth and to Anthony Whitworth for the stimulating discussions. 
Part of the activities relating to this work were supported by a China-Cardiff Centre Competition Fund, from Cardiff University.
We are grateful to the University of Central Lancashire for support provided through the Distinguished Visitors Programme.
M.B.N.K. was supported by the Peter and Patricia Gruber Foundation through the PPGF fellowship, by the Peking University One Hundred Talent Fund (985), and by the National Natural Science Foundation of China (grants 11010237, 11050110414, 11173004). This publication was made possible through the support of a grant from the John Templeton Foundation and National Astronomical Observatories of Chinese Academy of Sciences. The opinions expressed in this publication are those of the author(s) do not necessarily reflect the views of the John Templeton Foundation or National Astronomical Observatories of Chinese Academy of Sciences. The funds from John Templeton Foundation were awarded in a grant to The University of Chicago which also managed the program in conjunction with National Astronomical Observatories, Chinese Academy of Sciences.

\end{document}